\documentclass[%
reprint,
amsmath,amssymb,
aps,
prl,
longbibliography,
superscriptaddress,
floatfix,
]{revtex4-1}
\usepackage[export]{adjustbox}
\usepackage[caption=false]{subfig}
\usepackage{amssymb}
\usepackage{amsmath}
\usepackage{color}
\usepackage{siunitx}
\usepackage{verbatim}
\usepackage{longtable}
\DeclareMathAlphabet\mathbfcal{OMS}{cmsy}{b}{n}

\usepackage{graphicx}
\usepackage{dcolumn}
\usepackage{bm}
\usepackage{hyperref}
\hypersetup{colorlinks=true,citecolor=magenta,urlcolor=cyan}

\usepackage{xcolor}

\begin{document}
	
\title{Electrification in granular gases leads to constrained fractal growth}
	
	\author{Chamkor Singh}
	\affiliation{Max Planck Institute for Dynamics and Self-Organization (MPIDS), 37077, G\"ottingen, Germany}%
	\affiliation{Georg-August-Universit\"at G\"ottingen, Friedrich-Hund-Platz 1, 37077 G\"ottingen, Germany}
	\author{Marco G. Mazza}%
	\affiliation{Max Planck Institute for Dynamics and Self-Organization (MPIDS), 37077, G\"ottingen, Germany}%
	\affiliation{Interdisciplinary Centre for Mathematical Modelling and Department of Mathematical Sciences, Loughborough University, Loughborough, Leicestershire LE11 3TU, United Kingdom}
	
	\date{\today}

\begin{abstract}
The empirical observation of aggregation of dielectric particles under the influence of electrostatic forces lies at the origin of the theory of electricity. The growth of clusters formed of small grains underpins a range of phenomena from the early stages of planetesimal formation to aerosols.
However, the collective effects of Coulomb forces on the nonequilibrium dynamics and aggregation process in a granular gas -- a model representative of the above physical processes -- have so far evaded theoretical scrutiny. 
Here, we establish a hydrodynamic description of aggregating granular gases that exchange charges upon collisions and interact via the long-ranged Coulomb forces. We analytically derive the governing equations for the evolution of granular temperature, charge variance, and number density for homogeneous and quasi-monodisperse aggregation. We find that, once the aggregates are formed, the system obeys a physical constraint of nearly constant dimensionless ratio of characteristic electrostatic to kinetic energy $\mathcal{B}(t)\le 1$. This constraint on the collective evolution of charged clusters is confirmed both by the theory and the detailed molecular dynamics simulations. The inhomogeneous aggregation of monomers and clusters in their mutual electrostatic field proceeds in a fractal manner. Our theoretical framework is extendable to more precise charge exchange mechanism, a current focus of extensive experimentation. Furthermore, it illustrates the collective role of long-ranged interactions in dissipative gases and can lead to novel designing principles in particulate systems.
\end{abstract}

\maketitle

\section{Introduction}
The electrostatic aggregation of small particles is ubiquitous in nature and ranks among the oldest scientific observations. Caused by collisional or frictional interactions among grains, large amounts of positive and negative charges can be generated. These clusters have far-reaching consequences: from aerosol formation to nanoparticle stabilization \cite{castellanos2005relationship,schwager2008fractal}, planetesimal formation, and the dynamics of the interstellar dust \cite{wesson1973accretion,harper2017electrification,brilliantov2015size,blum2006dust}. The processes accompanying granular collisions, charge buildup and subsequent charge separation can also lead to catastrophic events such as silo failure, or dust explosions.

Experimental investigations of the effects of tribocharging date back to Faraday, and recent \emph{in situ} investigations have revealed important results \cite{jungmann2018sticking,lee2015direct,yoshimatsu2017self,poppe2000experiments,haeberle2018double}. However, technical difficulties
plague even careful experiments and  often impede their unambiguous interpretation \cite{spahn2015granular}. A source of these difficulties is the lack of consensus about whether electrostatics facilitate or hinder the aggregation process of a large collection of granular particles \cite{spahn2015granular}. Despite considerable effort  \cite{ivlev2002coagulation,dammer2004self,muller2011homogeneous,ulrich2009cooling,brilliantov2018increasing,liu2018cluster,takada2017homogeneous,shinbrot2018parlad}
a statistico-mechanical description of aggregation in a dissipative granular system with a mechanism of charge transfer is still lacking. The theoretical treatment requires reconsideration of the dissipation of kinetic energy conventionally described by a monotonic dependence of the coefficient of restitution on velocities $\epsilon(v)$, and also inclusion of long-range electrostatic forces due to the dynamically-changing charge production. Understanding the growth of charged aggregates requires a statistical approach due to the different kinetic properties and aggregate morphology.

In this work, we present a modified Boltzmann description for the inelastic and aggregative collisions of grains that interact via Coulomb forces, and exchange charges upon collision. We derive the hydrodynamic equations for the number density $n$, the granular temperature $T$, and the charge fluctuations $\langle \delta q^2 \rangle$ of the aggregates under the assumptions of homogeneous and quasi-monodisperse aggregation. We find that the dynamics of the charged granular gas approach, but do not overcome, a limiting behavior marked by the value of the dimensionless ratio of characteristic electrostatic to thermal energy. 
	\begin{figure*}[t!]
		\includegraphics[width=0.95\linewidth]{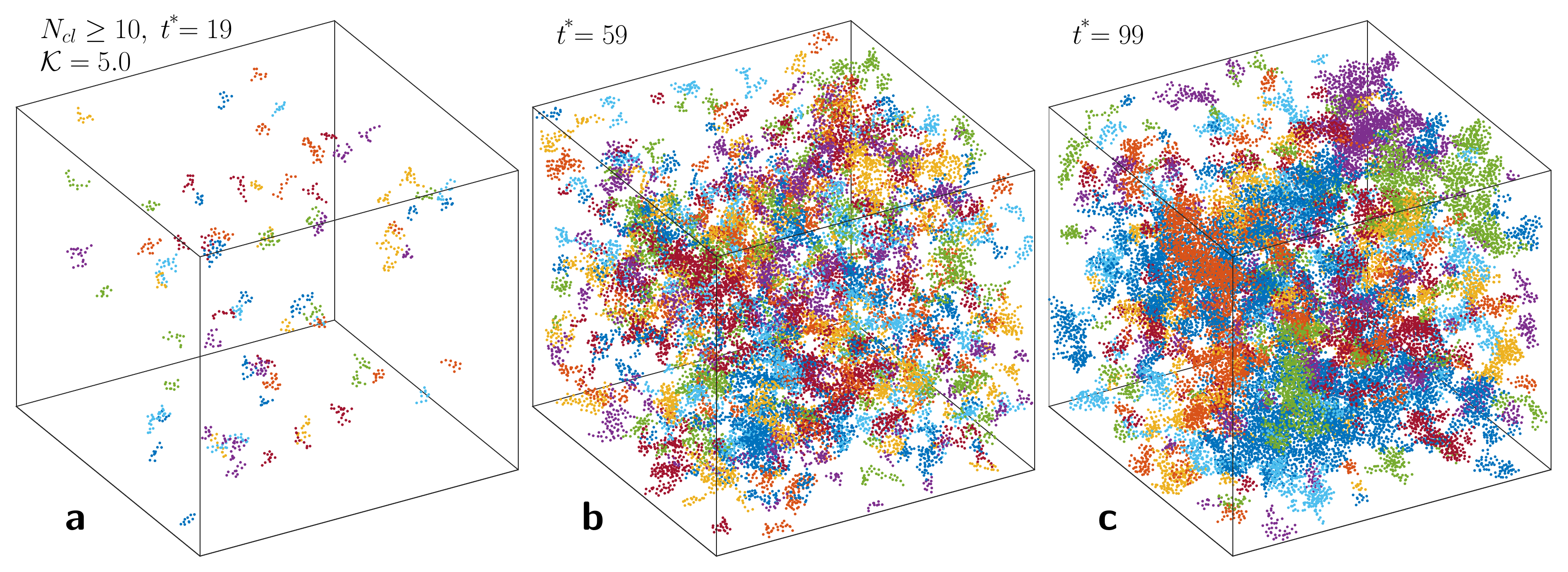}
 		\caption{Snapshots of the aggregating charged granular gas at different non-dimensional times: ({a}) $t^*=19$, ({b}) $t^*=59$, and (c) $t^*=99$. Here clusters containing 10 particles or more ($N_{cl}\geq 10$) are shown, and each color represents a different cluster.
        Clusters are identified on the basis of the monomer distances: if the centers of two particles are separated by a particle diameter, or less, they belong to the same cluster. See supplementary for non-dimensional time $t^*=t v_\textrm{ref}/d_0$, Coulomb force strength $\mathcal{K}$, and other reference scales.}
		\label{clusters}
	\end{figure*}

To bolster our results, and explicitly consider fluctuations in dynamics  and morphological structures, we also use three-dimensional molecular dynamics (MD) simulations that explicitly include Coulombic interactions and a charge-exchange mechanism. We find that the granular dynamics agree quantitatively with the predictions of the Boltzmann equation. The cooling gas undergoes a transition from a dissipative to an aggregative phase marked by a crossover in the advective transport. We explore the morphological dynamics of the inhomogeneous aggregation via the mean fractal dimension and their interplay with the macroscopic flow. 

\section{Kinetics}
In general, agglomeration in a three-dimensional collisionally charging cooling granular gas is a spatially inhomogeneous process which involves the interplay between dissipation, time-varying size distribution of aggregates, charge fluctuations and exchange mechanism during collisions, long-range forces, and collective effects~\cite{singh2018early}. This complexity is illustrated in Fig.~\ref{clusters} which shows snapshots of cooling clusters from a typical MD simulation, beginning from a homogeneous and neutral state (see supplementary for MD). In the following we establish a modified Boltzmann approach for this intricate dynamics of aggregation process, which predicts novel physical limits.

\begin{figure}[t!]
	\includegraphics[width=0.9\columnwidth]{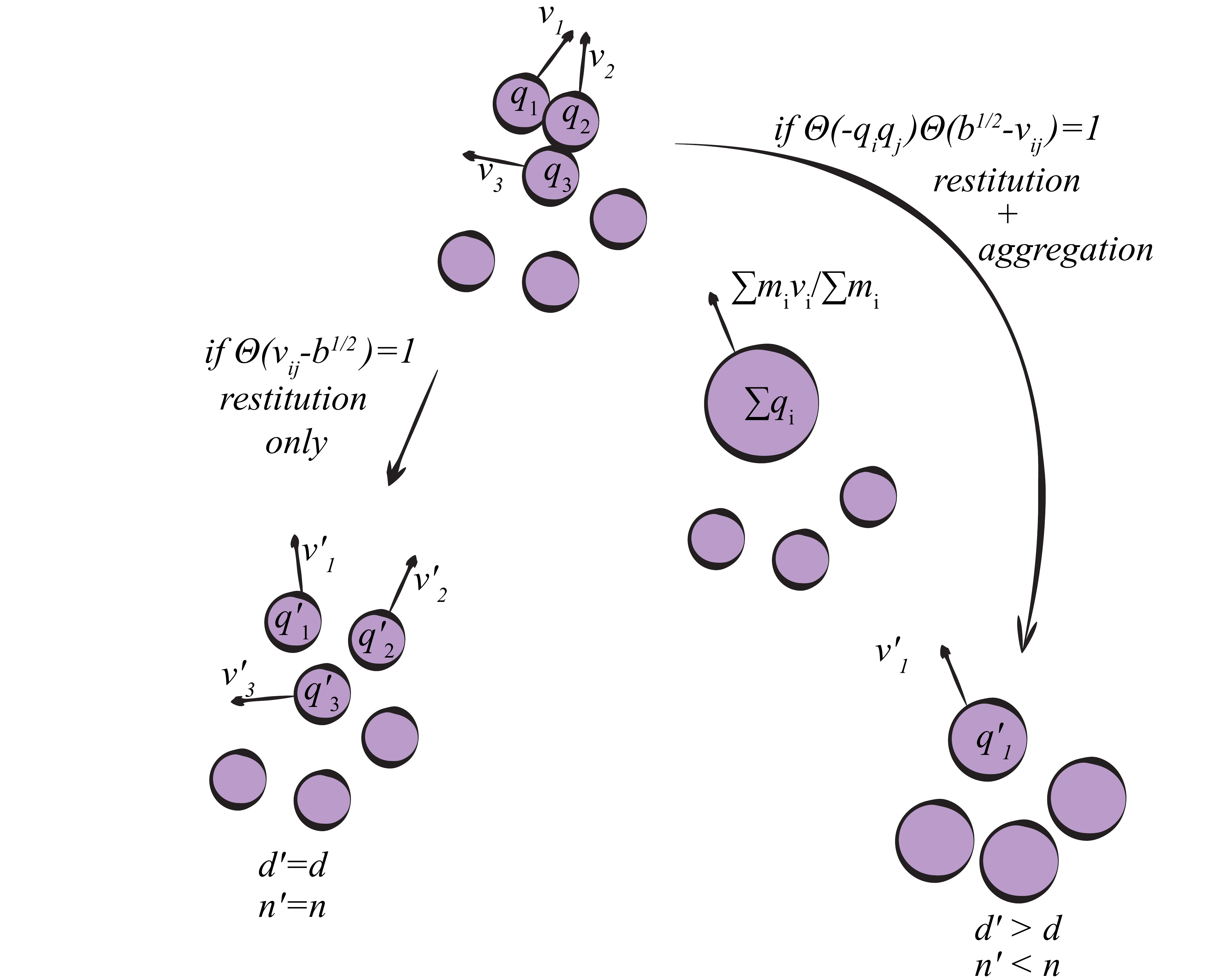}
	\caption{When particles collide two pathways are considered in the theory depending on their velocities and charges: (\emph{i}) a typical inelastic collision (dissipation only) with charge exchange, or (\emph{ii}) an inelastic collision that leads to the aggregation and merging of charges. Primed variables represent post-collision or post-aggregation values. 
}\label{schematic}
\end{figure}

We consider the single particle probability distribution function $f=f(\mathbf{r}, t; \mathbf{v},q,d)$, where the particle velocity $\mathbf{v}$, charge $q$, position $\mathbf{r}$ and particle size $d$, are the phase space variables, and $t$ denotes time. We specialize to a homogeneous and quasi-monodisperse aggregation scenario ({\it i.e.} the size is assumed to vary in time but spatially mono-dispersed, see schematic representation in Fig.~\ref{schematic}). Under these limits, the spatial and particle-size dependence of $f$ drops out, {\it i.e.} $f=f(t;\mathbf{v},q)$ only, and its time evolution is given by the simplified Boltzmann equation \cite{pitaevskii2012physical,brilliantov2010kinetic}	
\begin{align}
    \frac{\partial f}{\partial t} = I_\mathrm{coll},
\end{align}
valid at any time instant $t$. Here we define $I_\mathrm{coll}$ as the modified collision integral which includes dissipation as well as charge exchange during particle collisions. We will now elucidate how the charge exchange mechanism and particle size growth modify the collisions.

Let us consider contact collisions of particles $i$ and $j$ with pre-collision velocity-charge values $(\mathbf{v}_i ,q_i)$ and $(\mathbf{v}_j ,q_j)$, respectively. In the ensemble picture, particle collisions will change $f(t;\mathbf{v},q)$ in the infinitesimal phase-space volumes $d\mathbf{v}_i dq_i$ and $d\mathbf{v}_j dq_j$, centered around $(\mathbf{v}_i ,q_i)$ and $(\mathbf{v}_j ,q_j)$, respectively. The number of direct collisions $N_c^-$ per unit spatial volume which lead to loss of particles from the intervals $d\mathbf{v}_i dq_i$ and $d\mathbf{v}_j dq_j$ in time $\Delta t$ are 	
	\begin{align}
	N_c^- = f_i d\mathbf{v}_i dq_i  f_j d\mathbf{v}_j dq_j |\mathbf{v}_{ij}\cdot \mathbf{n}| \Theta (-\mathbf{v}_{ij}\cdot \mathbf{n})   \Theta_q d\sigma\Delta t\,,
	\end{align}
where $\mathbf{v}_{ij}\equiv\mathbf{v}_{i}-\mathbf{v}_{j}$, $\mathbf{n}$ is the unit vector at collision pointing from the center of particle $i$ towards particle $j$, and $d\sigma$ is the differential collision cross-section. The Heaviside step function $\Theta (-\mathbf{v}_{ij}\cdot \mathbf{n})$ selects particles coming towards $i$, while we use $\Theta_q \equiv \Theta\left(\frac{1}{2}mv_{ij}^2 - \frac{k_e q_i q_j}{d}\right)$ to ensure that a contact with an approaching particle takes place only when the Coulomb energy barrier can be overcome, where $k_e=1/(4\pi\epsilon_0)$, $\epsilon_0=8.854\times 10^{-12}$ F m$^{-1}$ is the vacuum permittivity, and $d$ is the particle diameter at time $t$. If the interaction is repulsive, ${k_e q_i q_j}/{d}$ is positive, and $\Theta_q=1$ only if $\frac{1}{2}mv_{ij}^2 > {k_e q_i q_j}/{d}$. In case of attractive interaction, ${k_e q_i q_j}/{d}$ is negative and thus $\Theta_q=1$ always. Essentially, $\Theta_q$ filters repulsive interactions which do not lead to a physical contact between particles. 
	
Consider now particles with initial velocity-charge values $(\mathbf{v}_i'', q_i'')$ and $(\mathbf{v}_j'', q_j'')$ in the intervals $d\mathbf{v}_i'' dq_i'',\;d\mathbf{v}_j'' dq_j''$.
The number of particles $N_c^+$ per unit volume which, post-collision, enter the interval $d\mathbf{v}_i dq_i$ and $d\mathbf{v}_j dq_j$ in time $\Delta t$ is
	\begin{align}
	N_c^+=f_i'' d\mathbf{v}_i'' dq_i''  f_j'' d\mathbf{v}_j'' dq_j''  |\mathbf{v}_{ij}''\cdot \mathbf{n}| \Theta (-\mathbf{v}_{ij}''\cdot \mathbf{n})   \Theta_q''d\sigma''\Delta t\,.
	\end{align}  
	The net change $\Delta N_c\equiv N_c^+ - N_c^-$ of number of particles in time $\Delta t$ per unit volume, then reads (see Supplementary)
	\begin{align}\nonumber
 \Delta N_c =&
	\left(\frac{1}{\epsilon(v_{ij})} \mathbf{J}^{v_{ij}}_{ij}  \mathbf{J}^{q}_{ij} f_i'' f_j'' - f_i f_j  \right) |\mathbf{v}_{ij}\cdot \mathbf{n}|\\
	& \times  \Theta (-\mathbf{v}_{ij}\cdot \mathbf{n})  d\mathbf{v}_j dq_j d\sigma \Theta_q\Delta t\,,
	\end{align}	
where $\mathbf{J}^{v_{ij}}_{ij}$ and $\mathbf{J}^{q}_{ij}$ are the Jacobians of the transformations for $d\mathbf{v}_i'' d\mathbf{v}_j'' \rightarrow d\mathbf{v}_i d\mathbf{v}_j$ and $dq_i'' dq_j'' \rightarrow dq_i dq_j$, respectively, which lump together the microscopic details of the collision process, namely dissipation and charge exchange in the present study. 

Integrating over all incoming particle velocities and charges  from all directions, dividing by $\Delta t$ and taking the limit $\Delta t \to 0$, we obtain the formal expression for the modified collision integral 	
	\begin{align} \nonumber
	I_\mathrm{coll} = &
	\int \left(\frac{1}{\epsilon(v_{ij})} \mathbf{J}^{v_{ij}}_{ij}  \mathbf{J}^{q}_{ij} f_i'' f_j'' - f_i f_j \right) |\mathbf{v}_{ij}\cdot \mathbf{n}|\\ \nonumber & \times \Theta (-\mathbf{v}_{ij}\cdot \mathbf{n})  d\mathbf{v}_j dq_j d\sigma \Theta_q.
	\end{align}	
Here we assume that the differential collision cross section and the contact condition specified by $\Theta_q$ retain their form for direct and inverse collisions. The particle encounters which do not lead to a physical contact have been excluded using $\Theta_q$. While taking moments of $I_\mathrm{coll}$, a fraction of those contact collisions that lead to aggregation is accounted for by taking the limit $\epsilon = 0$ for certain conditions on the relative velocity $v_{ij}$, and by considering the charge transferred to particle $i$ equal to the charge on particle $j$ (see Supplementary). In $I_\mathrm{coll}$, distant encounters, which do not lead to a contact between particles (glancing collisions) are neglected and the charge exchange and dissipation is considered only during the contact. The long-range effect is incorporated via the collision cross-section.

After setting up the collision integral, we derive the macroscopic changes of number density $n$, granular temperature $T$, and the charge variance $\langle\delta q^2\rangle$ by taking the moments of the Boltzmann equation (see Supplementary). The particles are initially neutral and the charge on them is altered either by collisions or during aggregation. 
However, due to charge conservation during collisions and aggregation, the system remains globally neutral and the mean charge variation $\langle \delta q \rangle$ is zero. 
The next  choice is thus $\langle\delta q^2\rangle$. In order to obtain closed form equations, and for analytical tractability, we assume quasi-monodispersity and homogeneity of the aggregating granular gas at any given time, as illustrated in Fig. \ref{schematic}. This means that during aggregation the mass of the clusters is assumed to grow homogeneously, while their numbers decrease in a given volume.
	
We assume that the charge and velocity distributions are uncorrelated, and their properly scaled form remains Gaussian (see Supplementary). After integration we find the governing equations 
    \begin{align}
	\frac{\partial n}{\partial t}&= -n^2 T^\frac{1}{2} g_1(\mathcal{B},C^n_{agg}),
	\label{eq_n}
	\end{align}
	\begin{align}
	&\frac{3}{2}\frac{\partial T}{\partial t}=
	-n^2 T^\frac{8}{5} g_2(\mathcal{B},C^T_{res})
	+n^2 T^\frac{3}{2} g_3(\mathcal{B},C^T_{agg}),
	\label{eq_T}
	\end{align}
	\begin{align}
	\frac{\partial \langle \delta q^2 \rangle}{\partial t}=
	n^2 T^{(\eta+\frac{1}{2})}g_4(\mathcal{B},C^q_{res})
	-n^2 \langle \delta q^2\rangle T^\frac{1}{2}g_5(\mathcal{B},C^q_{agg}),
	\label{eq_qq}
	\end{align}
which are coupled via a time-dependent dimensionless ratio
\begin{align}
\mathcal{B}(t)\equiv k_e \frac{\langle \delta q^2 \rangle(t)} {T(t)d(t)},
\label{eq_bjerrum}
\end{align}
between charge variance,  granular temperature, and  aggregate size. The terms $g_k$ are time-dependent functions of $\mathcal{B}$ and material constants $C_{res}, \;C_{agg}$ (Supplementary Table~I). We term the ratio $\mathcal{B}$ as Bjerrum number. In Eq.~\eqref{eq_bjerrum}, $d$ represents the size of a particle, also evolving with time during aggregation [Fig.~\ref{schematic}]. Notice that as $f$ is considered independent of $d$ during aggregation, an explicit equation for $d$ is required. For this we consider the total mass $M$, system volume $V$, and particle material density $\rho_p$ to be constant, which fixes the relation between particle size $d$ and particle number density $n$, according to 
\begin{align}
d(t) = \left[\frac{6M}{\pi n(t) V\rho_p}\right]^{\frac{1}{3}},
\label{eq_d}
\end{align}
and closes the equation set~\eqref{eq_n}-\eqref{eq_qq}. The above set of equations is consistent with the classical Haff's law in the absence of collisional charging. In this limit $\langle \delta q^2 \rangle=0$, $\mathcal{B}=0$, and we obtain $\frac{\partial n}{\partial t}=0$, $\frac{\partial \langle \delta q^2 \rangle}{\partial t}=0$ and $\frac{3}{2}\frac{\partial T}{\partial t}=
-T^{8/5} \left[\frac{\pi C^T_{res}}{2}\right]$, whose solution is Haff's law.  
\begin{figure}
	\includegraphics[width=1.0\linewidth]{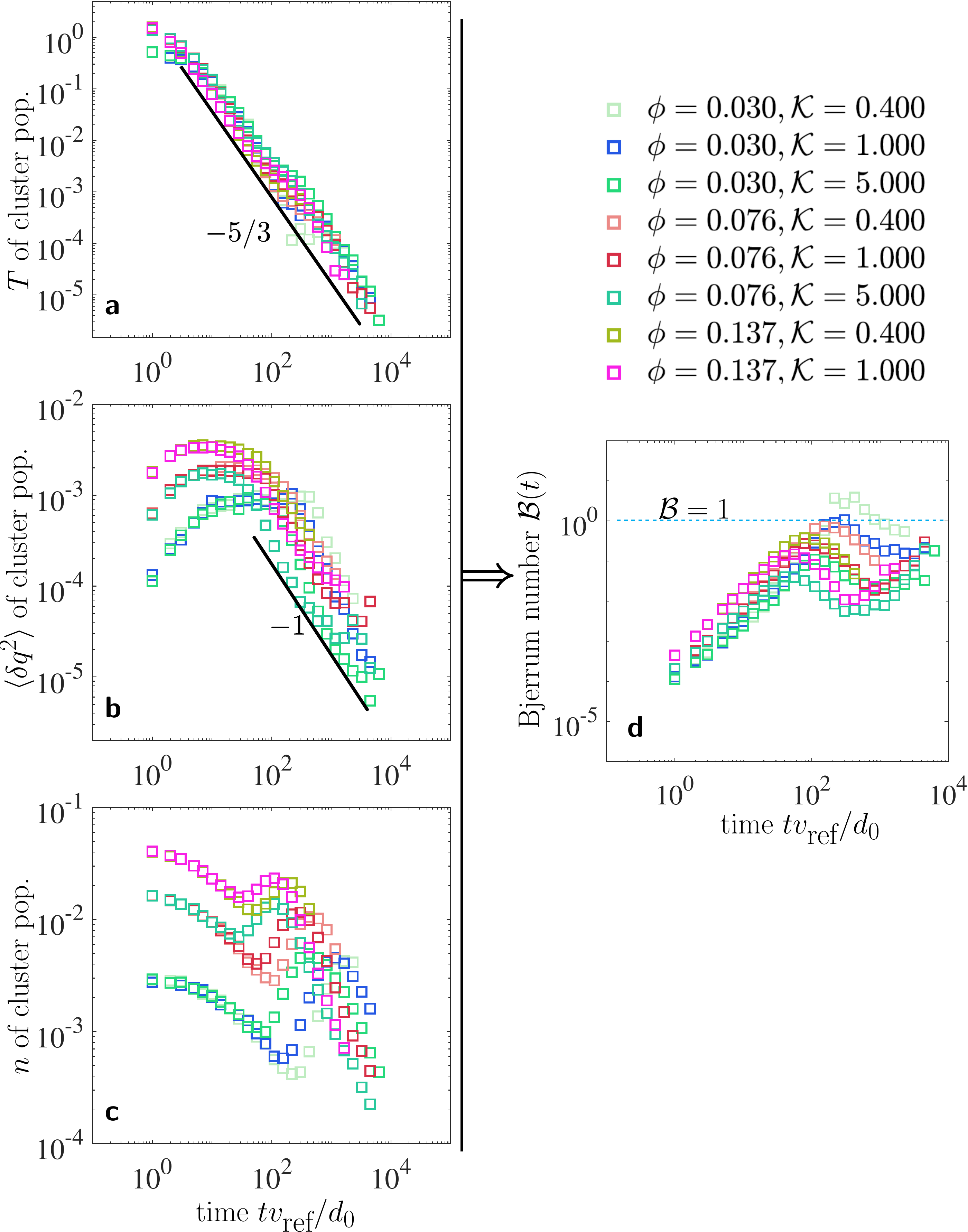}
	\caption{Evolution of temperature (a) $T$ of cluster population, 
    (b) charge variance $\langle\delta q^2\rangle$ of cluster population, and (c) number density $n$ of cluster population, for different monomer filling fractions $\phi$ and charge strength $\mathcal{K}$. (d) The granular temperature, charge variance and average size of the cluster population during aggregation evolve in such a manner that their non-dimensional combination $\mathcal{B}(t)=k_e\langle \delta q^2\rangle/(Td)  \le 1$ (see also Fig.~\ref{fig_R}). Both temperature and charge variance of cluster population decay as power laws. The number density evolution, however, is highly dynamic and exhibits a non-monotonic behavior due to emergence of macroscopic flow (see Supplementary for $\phi$, $\mathcal{K}$, $v_\text{ref}$, $d_0$ and macroscopic flow).}
	\label{fig_T_q_K}
\end{figure}

\section{Results and Discussion}
The time evolution of $T$, $\langle \delta q^2 \rangle$ and $n$ of aggregate population from MD simulations is shown in Fig.~\ref{fig_T_q_K}. The aggregate temperature from simulations is extracted as $\frac{3}{2}T=\frac{1}{N_\mathrm{agg}}\sum_i \frac{1}{2} m_i [(\mathbf{v}_i-\mathbf{V})^2]$. Notice that $\mathbf{v}_i$ and $m_i$ are center of mass velocity, and mass of the $i^{th}$ aggregate respectively, and should not be confused with monomer velocities and masses. $\mathbf{V}$ is the local advective velocity  in the neighborhood of $i^{th}$ aggregate. Similarly, $\langle \delta q^2\rangle=\frac{1}{N_\mathrm{agg}}\sum_i (q_i-\langle q\rangle)^2$, and $n=N_\mathrm{agg}/V$, where $N_\mathrm{agg}$ is the total number of aggregates and $V$ is the system volume. 

Initially ($tv_\textrm{ref}/l_\textrm{ref} < 10^2$), the relative collision velocities $v_{ij}$ remain larger than the time varying threshold $\sqrt{b}\equiv\sqrt{\frac{k_e |q_i q_j|}{2Td}}$ (see Supplementary for the treatment of threshold $b$). 
In this time regime, the collisions are primarily restitutive, leading to either Coulomb scattering without collision, or charge exchange and dissipation without considerable aggregation. The dissipation reduces $T$ [Fig.~\ref{fig_T_q_K}(a)] while the charge exchange increases $\langle \delta q^2 \rangle$ [Fig.~\ref{fig_T_q_K}(b)]. In this time regime the number density $n$, and thus size $d$ of the aggregates, is altered only moderately due to those low velocity attractive monomer encounters which lead to aggregation [Fig.~\ref{fig_T_q_K}(c)]. 

As a result of our kinetic formulation, the dynamics of $n$, $T$ and $\langle \delta q^2 \rangle$ can be collected into evolution of $\mathcal{B}$, shown in Fig.~\ref{fig_T_q_K}(d). The Bjerrum number $\mathcal{B}$ initially increases, which indicates that temperature decreases at a faster rate than the rate of increase of charge variance and the aggregate size. As the relative velocities $v_{ij}$ approach the threshold $\sqrt{b}$, $\mathcal{B}\to 1$. Near this time, the dynamics cross over to aggregative collapse. The individual particles, or monomers, cluster in such a way that the charge variance of the cluster population now begins to reduce. The temperature of the aggregate population keeps decreasing at the same rate with a slight dip near the aggregative collapse. The number density starts to evolve non-monotonically. We explore the non-monotonicity in next section and in the Supplementary. These results are robust under  variation of the initial monomer filling fraction $\phi$ and the charge strength $\mathcal{K}$ [Fig.~\ref{fig_T_q_K}].

\begin{figure}[t!]\centering
	\includegraphics[width=0.8\columnwidth]{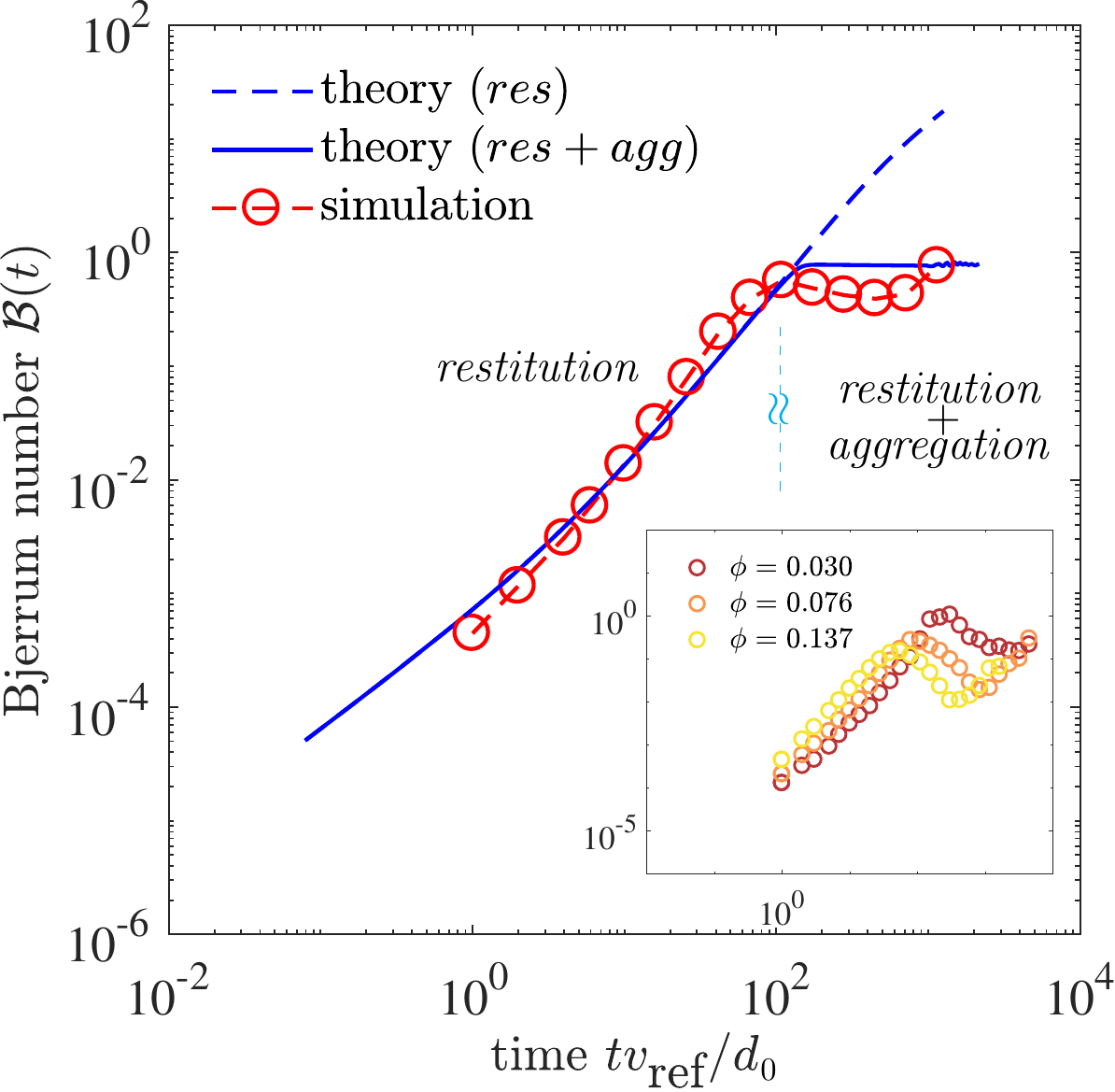}
	\caption{The granular temperature, charge variance and average size of the cluster population during aggregation evolve in such a manner that their non-dimensional combination $\mathcal{B}(t)=k_e\langle \delta q^2\rangle/(Td) \le 1$. This is not captured in the kinetic theory if only restitutive (no aggregation) collisions are considered. The granular MD simulations confirm the analytical results. (inset for different monomer filling fraction $\phi$). }
	\label{fig_R}
\end{figure}

After the initial time regime and the aggregative collapse, the crossover in the dynamics is depicted in Fig.~\ref{fig_R}, where the evolution of the combination $\mathcal{B}$, and its comparison with the solution of Eq.~\eqref{eq_n}-\eqref{eq_qq} is highlighted. We solve the full kinetic equations  Eq.~\eqref{eq_n}-\eqref{eq_qq} including the  aggregation kinetics [Fig.~\ref{fig_R} (solid line)]. We also solve Eq.~\eqref{eq_n}-\eqref{eq_qq} for a system with only dissipative collisions and  without aggregation; hence, the cluster size $d$ remains unchanged. These results are shown in Fig.~\ref{fig_R} (dashed line). In this limit of only restitutive kinetics, $\mathcal{B}$ increases continuously above the limiting value $1$. The purely restitutive kinetic theory thus fails to predict the MD results. When aggregation is explicitly treated (solid line), the theory predicts an upper limit during the growth. The theory shows that once aggregation sets in, the aggregating granular gas obeys the constraint
\begin{align}
\mathcal{B}(t)\le 1\,.
\label{eq_constraint}
\end{align}

The upper physical limit predicted in the theory, $\mathcal{B}(t)< 1$, is endorsed by the granular MD simulations under moderate variation of $\phi$ and $\mathcal{K}$. It is notable that at later times, the limit $\mathcal{B}\leq1$ allows the right hand sides of the equations for number density, temperature and charge variance [Eq.~\eqref{eq_n}-\eqref{eq_qq}] to remain real valued during the aggregation process. This mathematical indication confirms the effectiveness of the quasi-monodisperse picture [Fig.~\ref{schematic}] considered in the present study.

The initiation of aggregation brings about a power law decay in the charge variance [Fig.~\ref{fig_T_q_K}(b)]. It is notable that a different charge exchange model might provide a different charge buildup rate during the purely dissipative (restitutive) phase. 
However, the decay of charge variance during the aggregative phase is not expected to be influenced by charge exchange mechanism. The reason is that aggregation sets in at relatively low temperature where the thermal motion of monomers, if any, inside the clusters is significantly decreased, and thus collisional charge-exchange is expected to be negligible. 
Also, in our kinetic theory description, the charge transferred to particle $i$ during aggregation with particle $j$ is equal to the charge on particle $j$, as they  merge into one single particle [Fig.~\ref{schematic}]. Thus, the charge variance of the cluster population is reduced by the aggregation process, rather than by the collisional charge exchange. 

Our key theoretical finding is that after the aggregative collapse, the decay of the charge variance of aggregate population and the growth of the size of the aggregates is balanced by the decay of temperature during the aggregation, resulting in the stationary value of $\mathcal{B}(t)$. The constraint $\mathcal{B}\le 1$  is robust in the theory, while the granular MD simulations suggest $\mathcal{B}(t)< 1$ and confirm the upper limit of $\mathcal{B}(t)$. It is also intriguing that the temperature of the cluster population still closely follows  Haff's law for different $\phi$ and $\mathcal{K}$, despite complex heterogeneous aggregation-fragmentation events and the long-range electrostatic interactions.

The number density's temporal evolution obtained from the MD simulations reveals a more intricate non-monotonic dynamics. It initially begins to decrease during small aggregate formations due to low velocity attractive monomer encounters. In an intermediate time regime, the emergence of macroscopic particle fluxes trigger fragmentation events and the aggregate numbers increase. We quantify the emergence of macroscopic flow using the Mach number (see Supplementary and Media therein). After this intermediate time regime, the aggregation again takes over and the number density of clusters starts to decrease.  The non-monotonic evolution of $n$ causes a dip in $\mathcal{B}(t)$ after the aggregative collapse ($tv_\mathrm{ref}/l_\mathrm{ref}>10^2$) [Fig.~\ref{fig_T_q_K} and \ref{fig_R}]. 
The distribution of clusters neglected in the homogeneous and quasi-monodisperse aggregation picture adds to the complexity of $\mathcal{B}$'s evolution. However, the theory clearly predicts the growth  of $\mathcal{B}$ and selects an upper limit. To further explore the mechanisms behind the non-monotonic evolution of $n$, we explore the spatially heterogeneous cluster dynamics and nature of the structures from the MD simulations.

\subsection{Inhomogeneous aggregation and fractal growth}
To gain access to the spatial structure formation in the gas,  we perform a detailed cluster analysis of the results from granular MD simulation, see Fig.~\ref{clusters}.
\begin{figure}
    \centering
	\includegraphics[width=0.98\linewidth]{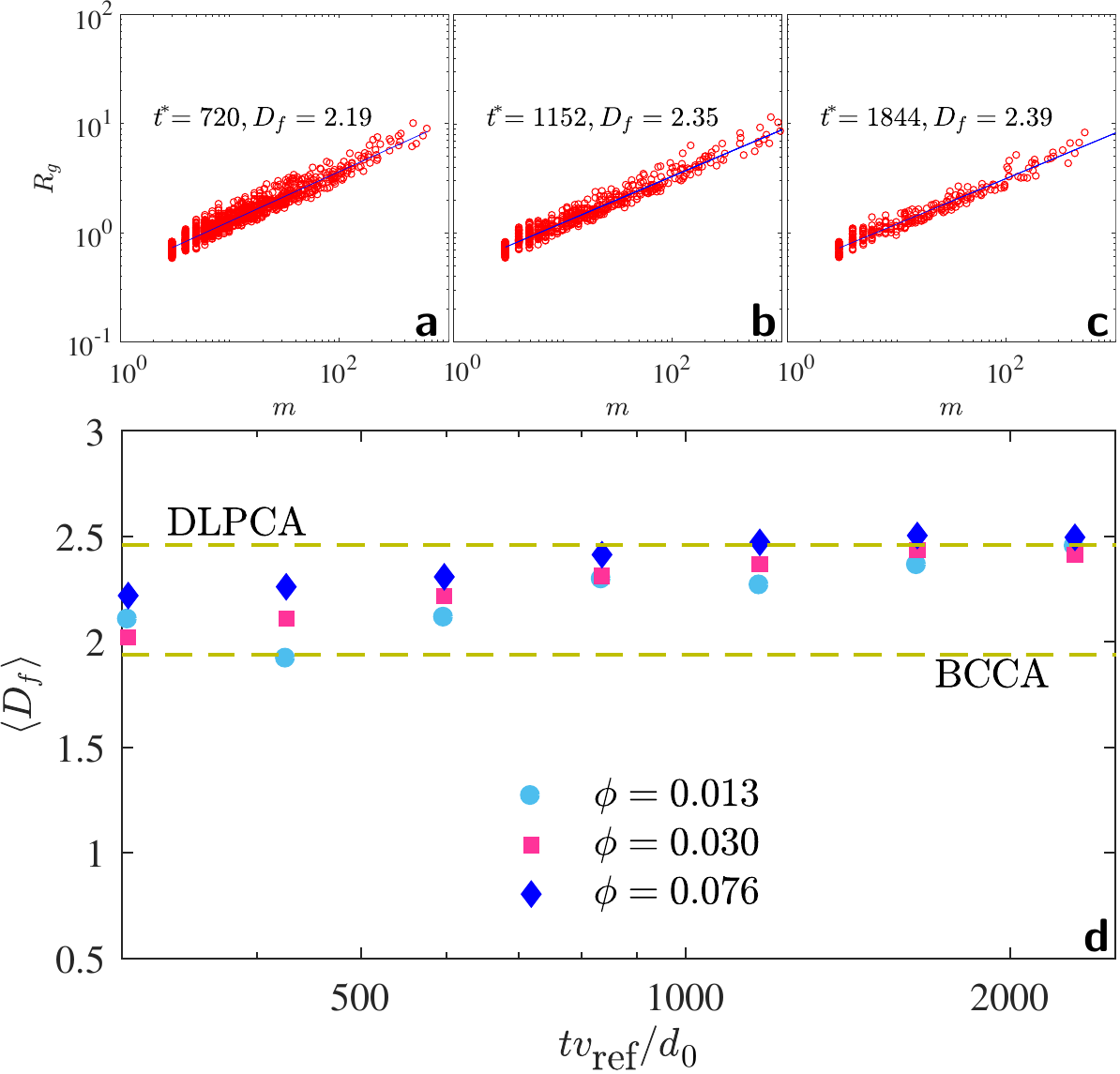}
	\caption{ ({a-c}) The scaling between cluster mass $m$ and their radius of gyration $R_g$, $m\sim R_g^{D_f}$ at different times, and ({d}). the time evolution of $\langle D_f \rangle$, thus obtained, for different filling fractions. The average fractal dimension in the aggregating charged gas varies across reported average values for ballistic cluster-cluster aggregation (BCCA, $\langle D_f \rangle \sim 1.94$) and diffusion-limited particle-cluster aggregation (DLPCA, $\langle D_f \rangle \sim 2.46$) \cite{blum2006dust,smirnov1990properties}. $t^*\equiv tv_\textrm{ref}/d_0$.}
	\label{Df}
\end{figure}
The morphology of the aggregates is studied by computing the average fractal dimension $\langle D_{f}\rangle$ \cite{mandelbrot1977fractals,jullien1987aggregation} of cluster population from the scaling relation $m \sim R_{g}^{\langle D_{f}\rangle}$ between cluster masses $m$, and radii of gyration $R_{g}=[\frac{1}{N_\textrm{mon}}\sum_i(\mathbf{r}_i - \mathbf{r}_\textrm{mean})^2]^{1/2}$, where the index $i$ runs over total number of monomers $N_\textrm{mon}$ in a given aggregate. Figure~\ref{Df}(a-c) show scatter plots for  $R_{g}$ versus $m$  at different times, and 
Fig.~\ref{Df}(d)  the time evolution of the exponent $\langle D_{f}\rangle$ for varying filling fraction. The average fractal dimension lies  between the  average values reported for the ballistic cluster-cluster aggregation (BCCA, $\langle D_f \rangle \simeq 1.94$) and the diffusion-limited particle-cluster aggregation (DLPCA, $\langle D_f \rangle \simeq 2.46$) \cite{blum2006dust,smirnov1990properties} models. In time, $\langle D_{f}\rangle$ is dynamic and changes across the two model limits. These results indicate that the aggregate structures retain their fractal nature over time. 

The BCCA and DLPCA are popular models for aggregation that have been used for neutral dust agglomeration (e.g hit and stick, ballistic motion, \cite{blum2006dust}), wet granulate aggregation (sticking due to capillary bridges and ballistic motion, \cite{ulrich2009cooling}), colloidal aggregation (van der Waals and repulsions \cite{lebovka2012aggregation}), and hit and stick agglomeration in Brownian particles under frictional drag \cite{kempf1999n}. The observation that $\langle D_f \rangle$ lies between the reported average values of $\langle D_f \rangle$ for BCCA and DLPCA indicates the presence of mixed characteristics from both of these simplified models. The size distribution in an aggregating, charged granular gas \cite{singh2018early} tends to resemble a DLPCA-like behavior where the smaller size aggregates are larger in number, in contrast to a BCCA-like model where the size distribution is typically bell-shaped \cite{blum2006dust}. On the other hand, the monomer motion is found to be highly sub-diffusive \cite{singh2018early} in agreement with the BCCA model. In addition, the Coulombic interactions will cause considerable deviations from the short-ranged or ballistic propagation typical of the BCCA or DLPCA models. We find that the long-range forces due to a bipolar charge distribution lead to the  value of $\langle D_{f}\rangle$ intermediate between  the above two aggregation models, indicated by dashed lines in Fig.~\ref{Df}(d). 

\subsection{Interplay between fractals and macroscopic flow}
Apart from the long-range effects, the morphology  of the aggregates is also altered by additional mechanisms. We discuss two physical processes that are not captured in the analytical theory, but that we investigate via our MD simulations. 

First, in our modified Boltzmann kinetic description, the collisions between aggregates at any given time are considered as  collisions between two spheres with sizes equal to the average size of the aggregate population. This is a quasi-monodisperse assumption typically used in cluster-cluster aggregation models. The quasi-monodispersity however neglects the morphology and surface irregularities of the colliding aggregates. Collisions between aggregates with large size difference, between aggregates and individual monomers, and the annihilation events are also simplified.  

Secondly, granular gases are characterized by the emergence of a convective flow \cite{hummel2016universal,brilliantov2010kinetic} which we find (see Supplementary and Media therein), in the present case, induces the non-monotonicity in the temporal evolution of the number density  [Fig.~\ref{fig_T_q_K}(c)]. 
Due to the macroscopic flow, aggregates which are weakly connected are prone to fragmentation. This results in an intermediate regime where the concentration of aggregates increases instead of decreasing.

Excluding the two above mechanisms explains the slight deviation of our quasi-monodisperse Boltzmann theory from the non-monotonic behavior of $\mathcal{B}(t)$ after the crossover to aggregative collapse.

\section{Conclusions}
We have derived the rate equations for the evolution of the number density $n$, granular temperature $T$, and charge variance $\langle \delta q^2 \rangle$ of the cluster population in a charged, aggregating granular gas. In contrast to well-known Smoluchowski-type equations, we have explicitly coupled $n$  to the decay of $T$  and charge variance. We have compared the results with three-dimensional molecular dynamics simulations and the outcomes of a detailed cluster analysis, and have explored the morphology of the aggregating structures via fractal dimension.

Taken together, our results indicate that the aggregation process in a charged granular gas is quite dynamic,  while respecting some physical constraints. The growth process obeys $\mathcal{B}(t)=k_e \langle \delta q^2 \rangle/(Td) \leq 1$, while morphologically, the clusters exhibit statistical self-similarity, persistent over time during the growth. The fractal dimension and growth of structures is intermediate between the BCCA and DLPCA models. We also demonstrate that the application of a purely dissipative kinetic treatment is not sufficient to make predictions about global observables such as $T$ and $\langle \delta q^2\rangle$ in an aggregating charged granular gas. 

Finally, we believe that our kinetic approach can be applied to study aggregation processes in systems such as wet granulates with ion transfer mechanism \cite{lee2018collisional,zhang2015electric}, dissipative cell or active particle collections under long-range hydrodynamic and electrostatic effects \cite{yan2016reconfiguring,friedl2009collective}, and charged ice-ice collisions \cite{dash2001theory}. 
\begin{widetext}
\section{Appendix\\ Kinetics and modified collision integral}
After obtaining number of direct collisions in Eq.~(2) in the main text, let us consider the number of particles $N_c^+$  per unit spatial volume having initial velocity-charge values $(\mathbf{v}_i'' ,q_i'')$ and $(\mathbf{v}_j'' ,q_j'')$ in the intervals $d\mathbf{v}_i'' dq_i''$  and $d\mathbf{v}_j'' dq_j''$ which, post-collision, enter the in the intervals $d\mathbf{v}_i dq_i$  and $d\mathbf{v}_jdq_j$ in time $\Delta t$
are	
\begin{align}
N_c^+=f_i'' d\mathbf{v}_i'' dq_i''  f_j'' d\mathbf{v}_j'' dq_j''  |\mathbf{v}_{ij}''\cdot \mathbf{n}| \Theta (-\mathbf{v}_{ij}''\cdot \mathbf{n})   \Theta_q''d\sigma''\Delta t\,,
\end{align}  
and thus the net increase of number of particles per unit time and volume is $N_c^+ - N_c^-$.
We can relate the primed velocities to the unprimed via 
\begin{align}
d\mathbf{v}_i'' d\mathbf{v}_j'' = \mathbf{J}^{v_{ij}}_{ij} d\mathbf{v}_i d\mathbf{v}_j,
\label{eq_jacobian_v}
\end{align}	
where $\mathbf{J}^{v_{ij}}_{ij}=1+(6/5)C_\epsilon v_{ij}^{1/5}+...$ is the Jacobian of the transformation for viscoelastic particles~\cite{brilliantov2010kinetic}. Here $C_\epsilon$ is a material constant. To obtain the transformation $d{q}_i'' d{q}_j''\rightarrow d{q}_i d{q}_j$, we consider the ratio of relative charges after and before the collision 
\begin{align}
r=\frac{q_i-q_j}{q_i''-q_j''},
\end{align}
and in addition we impose charge conservation during collisions
\begin{align}
{q_i+q_j}={q_i''+q_j''}.
\end{align}
The above two relations finally provide the transformation 
\begin{align}
d{q}_i'' d{q}_j''=\mathbf{J}^{q}_{ij}d{q}_i d{q}_j,
\label{eq_jacobian_q}
\end{align}
where, for example, $\mathbf{J}^{q}_{ij}=\frac{2}{r}$ for a constant $r$. This means that the differential charge-space volume element shrinks or expands by a factor of $r/2$. In general, for velocity and particle pre-charge dependent charge transfer, the expressions of $r$ and $\mathbf{J}^{q}_{ij}$ can be quite complicated as it depends on how the charge exchange takes place during collisions and its dependence on myriad factors (such as size, composition, and crystalline properties). Incorporating the above phase-space volume transformations due to collisions, the net change $\Delta N_c$ of number of particles per unit phase-space volume and in time $\Delta t$ reads
\begin{align}\nonumber
\Delta N_c =&
\left(\frac{1}{\epsilon(v_{ij})} \mathbf{J}^{v_{ij}}_{ij}  \mathbf{J}^{q}_{ij} f_i'' f_j'' - f_i f_j  \right) |\mathbf{v}_{ij}\cdot \mathbf{n}|
\Theta (-\mathbf{v}_{ij}\cdot \mathbf{n})  d\mathbf{v}_j dq_j d\sigma \Theta_q\Delta t\,,
\end{align}	
where we assume that the differential cross-section and the contact condition  specified by $\Theta_q$ are the same for direct and inverse collisions. Finally, dividing by $\Delta t$, and integrating over all incoming particle velocities and charges  from all directions in the limit $\Delta t \to 0$, we obtain a formal expression for the collision integral 	
\begin{align}
I_\mathrm{coll} = &
\int \left(\frac{1}{\epsilon(v_{ij})} \mathbf{J}^{v_{ij}}_{ij}  \mathbf{J}^{q}_{ij} f_i'' f_j'' - f_i f_j \right) |\mathbf{v}_{ij}\cdot \mathbf{n}|\Theta (-\mathbf{v}_{ij}\cdot \mathbf{n})  d\mathbf{v}_j dq_j d\sigma \Theta_q.
\end{align}	
At this point the particle encounters which do not lead to a physical contact have been excluded using $\Theta_q$, however, collisions that lead to aggregation have not been explicitly accounted. We do so by taking the limit $\epsilon = 0$ for certain conditions on the relative velocity $v_{ij}$, and by considering the charge transferred to particle $i$ equal to the charge on particle $j$ [Eq.~\eqref{eq_delta_psi_n}-\eqref{eq_delta_psi_q_agg} below]. In $I_\mathrm{coll}$, distant encounters, which do not lead to a contact between particles (glancing collisions) are neglected and the charge exchange and dissipation is considered only during the contact. The long-range effect is incorporated via collision cross-section.

\section{Splitting restitution and aggregation}
The time rate of change of the average of a microscopic quantity $\psi(\mathbf{v}_i,q_i)$ is obtained by multiplying the Boltzmann equation for $f_i$ by $\psi_i$ and integrating over $\mathbf{v}_i,q_i$, {\it i.e.}
\begin{align}
\frac{\partial \langle\psi \rangle}{\partial t} = \int d\mathbf{v}_i dq_i\psi_i\frac{\partial f_i}{\partial t} = \int d\mathbf{v}_i dq_i \psi_i I_\mathrm{coll}.
\end{align}
It can be shown that
\begin{align} 
\frac{\partial \langle\psi \rangle}{\partial t} &= \int d\mathbf{v}_i dq_i \psi_i I_\mathrm{coll}\nonumber \\
&=\frac{1}{2}\int d\mathbf{v}_i
d\mathbf{v}_j
d{q}_i
d{q}_j 
d\sigma
f_i f_j 
|\mathbf{v}_{ij}\cdot \mathbf{n}|
\Theta(-\mathbf{v}_{ij}\cdot \mathbf{n}) \Theta_q \Delta[\psi_i+\psi_j]\nonumber \\
&=\int d\mathbf{v}_i
d\mathbf{v}_j
d{q}_i
d{q}_j 
d\sigma
f_i f_j 
|\mathbf{v}_{ij}\cdot \mathbf{n}|
\Theta(-\mathbf{v}_{ij}\cdot \mathbf{n}) \Theta_q \Delta[\psi_i],\label{eq_avg_rate_of_change}
\end{align}
where $\Delta [\psi_i+\psi_j]=(\psi_i'+\psi_j'-\psi_i-\psi_j)$  and $\Delta [\psi_i]=(\psi_i'-\psi_i)$ is the change of $\psi$ during the collision between pair $i,j$, and the prime denotes a post collision value. We note that the transformations in Eq.~\eqref{eq_jacobian_v} and \eqref{eq_jacobian_q} are reversed back while integrating $I_\mathrm{coll}$ weighted with quantity of interest $\psi$. We consider the number density, the kinetic energy or granular temperature, and the charge variance (the system is globally neutral and the mean charge variation $\langle \delta q \rangle$ is zero), respectively 
\begin{align} 
(i)\;\psi&= n,\label{eq_psi_n}\\
(ii)\;\psi&=\frac{1}{2} m v^2,\label{eq_psi_T}\\
(iii)\;\psi&= (\delta q)^2 = (q-q_0)^2 = \left(q-\frac{q_i+q_j}{2}\right)^2.\label{eq_psi_q}
\end{align}
At this point we differentiate the restitutive or dissipative collisions from aggregative ones  by splitting $\Theta_q\Delta [\psi_i]$ as
\begin{align}\nonumber
\Theta_q \Delta [\psi_i] &= \Delta^{res} [\psi_i]\Theta\left(v_{ij}-\sqrt{b}\right),\\
&+ \Delta^{agg} [\psi_i] \Theta(-q_iq_j)\Theta\left(\sqrt{b}-v_{ij}\right),
\label{eq_psi_splitting}
\end{align}
where 
\begin{align}       \sqrt{b}\equiv\sqrt{\frac{2 k_e |q_i q_j|}{md}}.
\end{align} 
If $v_{ij}>\sqrt{b}$, the particles collide and separate after the collision irrespective of the sign of $q_iq_j$ (attractive or repulsive). This leads to dissipation of energy with finite non-zero $\epsilon=\epsilon(v_{ij})$, and charge exchange according to a specified rule. The aggregative part is zero in this case. If $v_{ij}<\sqrt{b}$ and $q_iq_j<0$ (attractive encounters at low velocities), the particles collide and aggregate with $\epsilon=0$, and with charge exchange to particle $i$ equal to $q_j$. If $v_{ij}<\sqrt{b}$ and $q_iq_j>0$ (repulsive encounters at low velocities), no physical contact takes place between the particles which leads to neither dissipation nor aggregation ($\Theta_q\Delta [\psi_i]=0$). Also represented schematically in Fig.~(2) in the main text.

The expressions for $\Delta^{res} [\psi_i]$ and $\Delta^{agg} [\psi_i]$ are obtained as follows. The particle number does not change during a dissipative collision but reduces by one in an aggregative collision, {\it i.e.}
\begin{align}\nonumber
\Delta^{res}_n[\psi_i+\psi_j]   &= 0, \\ 
\Delta^{agg}_n[\psi_i+\psi_j]_n &= -1.
\label{eq_delta_psi_n}
\end{align}
For the granular temperature,  
\begin{align}\nonumber
\Delta^{res}_T[\psi_i+\psi_j] &= -\frac{1}{2}m(1-\epsilon^2)(\mathbf{v}_{ij}\cdot \mathbf{n})^2,\\
\Delta^{agg}_T[\psi_i+\psi_j] &= -\frac{1}{2}m(\mathbf{v}_{ij}\cdot \mathbf{n})^2,
\label{eq_delta_psi_T}
\end{align}
where we take the limit $\epsilon = 0$ for the aggregation. 
The change in the charge variance is obtained as
\begin{align}\nonumber
\Delta^{res}_q[\psi_i] &= (\delta q_i^2)'- (\delta q_i^2)\\
&=(q_i'-q_i)^2+2(q_i'-q_i)(q_i-q_0),
\label{eq_delta_psi_charge}
\end{align}	
where $(q_i'-q_i)$ equals the charge transferred to particle $i$ during its collision with particle $j$, and $q_0=\frac{q_i+q_j}{2}$ is the mean charge on the pair. For the charge transfer, based on seminal experiments~\cite{blum2006dust}, we consider
\begin{align}
(q_i'-q_i)=C_{\Delta q} |\mathbf{v}_{ij}\cdot \mathbf{n}|^\eta \frac{q_i-q_0}{|q_i-q_0|},
\label{eq_dq}
\end{align}	
which is also obtainable if charge transferred is considered proportional to the contact area during the course of collisions. Using Eq.~\eqref{eq_dq} in Eq.~\eqref{eq_delta_psi_charge}, we get
\begin{align}
\Delta^{res}_q[\psi_i] = C_{\Delta q}^2 |\mathbf{v}_{ij}\cdot \mathbf{n}|^{2 \eta} 
+ 2C_{\Delta q}|\mathbf{v}_{ij}\cdot \mathbf{n}|^\eta (q_i-q_0).
\label{eq_delta_psi_q_res}	                        
\end{align}	
For aggregation, the charge tranferred to particle $i$ equals the charge on the merging particle $j$, {\it i.e}, $q_i'-q_i = q_j$, which gives
\begin{align}\nonumber
\Delta^{agg}_q[\psi_i] &= (\delta q_i^2)'- (\delta q_i^2)\\
&=q_iq_j.
\label{eq_delta_psi_q_agg}
\end{align}	

Putting Eq.~\eqref{eq_delta_psi_n},\eqref{eq_delta_psi_T},\eqref{eq_delta_psi_q_res}, and \eqref{eq_delta_psi_q_agg} in \eqref{eq_psi_splitting}, and then \eqref{eq_psi_splitting} in \eqref{eq_avg_rate_of_change}, the resulting integrals are solved, assuming the statistical independence of charge-velocity distribution function, {\it i.e.}, $f(\mathbf{v},q)=f(\mathbf{v})f(q)$, and assuming that their scaled form remains Gaussian. In addition to the charge exchange, the coefficient of restitution is taken as velocity dependent, {\it i.e}
\begin{align}
\epsilon = \epsilon(|\mathbf{v}_{ij}\cdot \mathbf{n}|) = 1 - C_\epsilon |\mathbf{v}_{ij}\cdot \mathbf{n}|^{1/5} + ...
\label{eq_coeff_of_rest}
\end{align}
while the long-range effects due to Coulomb interactions are taken into account by the change in collision cross section. After integration we obtain Eq.~(5)-(7) in the main text. The functions $g_k$ in Eq.~(5)-(7) have the forms
	\begin{align}
	g_1(\mathcal{B})&=
	\left[\frac{C^n_{agg}}{l_1^3l^3}\right]
	\left[ a^n_{1}\tan^{-1}\frac{l_1}{\mathcal{B}} + a^n_{2} +  a^n_{3}\tan^{-1}\frac{\mathcal{B}}{l} 
	+ a^n_{4}\tan^{-1}\frac{l_1}{\mathcal{B}} + a^n_{5} +  a^n_{6}\tan^{-1} 
	\frac{l}{\mathcal{B}}
	\right],\\
	g_2(\mathcal{B})&=
	\left[\frac{C^T_{res}}{l^5}\right] \left[a^T_1 + a^T_2\tan^{-1}\frac{\mathcal{B}}{l}\right],\\
	g_3(\mathcal{B})&=
	\left[\frac{C^T_{agg}}{l^5}\right]
	\left[
	a^T_3 + a^T_4 \cot^{-1}\frac{\mathcal{B}}{l}
	+a^T_5 + a^T_6 \tan^{-1}\frac{\mathcal{B}}{l} \right],\\
	g_4(\mathcal{B})&=
	\left[\frac{C^q_{res}}{l^5}\right] 
	\left[a^q_1 + a^q_2\tan^{-1}\frac{\mathcal{B}}{l}\right],\\
	g_5(\mathcal{B})&=C^q_{agg}
	\left[
	\frac{1}{a^q_3}
	\left(
	a^q_4 + a^q_5 \tan^{-1} \frac{\mathcal{B}}{l_1} + a^q_6 \tan^{-1} \frac{l}{\mathcal{B}} 
	\right)
	+\frac{1}{a^q_7}
	\left(
	a^q_8 \tan^{-1} \frac{l_1}{\mathcal{B}} + a^q_9 + a^q_{10} \tan^{-1} \frac{l}{\mathcal{B}}
	\right)
	\right],
	\label{eq_g5}
	\end{align}
where the coefficients $l$, $l_1$, $a^T_k$, $a^q_k$, $a^n_k$ are functions of $\mathcal{B}(t)$ [Table \ref{table_coeffs}].

\section{Derivation of the hydrodynamic equations (5)-(7)}

To solve the integrals in Eq.~\eqref{eq_avg_rate_of_change}
for different $\Delta [\psi_i]$ from
Eq.~\eqref{eq_delta_psi_n}-\eqref{eq_delta_psi_q_agg}, 
we assume that the normalized velocity as well as charge distribution of the aggregating particle population at any time remain Gaussian, and the two are uncorrelated, {\it i.e}
\begin{align}
f(\mathbf{v},q)=f(\mathbf{v})f(q)=n\left(\frac{m}{{2 \pi T}}\right)^\frac{3}{2} e^{-m v^2 / (2 T)} \left( \frac{1}{{2\pi \langle \delta q^2 \rangle}}\right)^\frac{1}{2}e^{-q^2 / (2 \langle \delta q^2 \rangle)}.
\label{eq_distributions}
\end{align}
In Fig.~\ref{fig_qdist}, we show charge distribution of monomers obtained from typical simulation runs, which is essentially the distribution before initiation of the aggregation. In Eq.~\eqref{eq_distributions} above, we assume that although granular temperature $T$ and charge variance $\langle\delta q^2\rangle$ do change with time due to restitution and aggregation, the shape of the scaled distribution remains close to Gaussian, and the increase of size/decrease of number of particles due to aggregation process does not alter scaled distribution shape.
\begin{figure}[h!]
	\centering
	\includegraphics[width=0.35\linewidth]{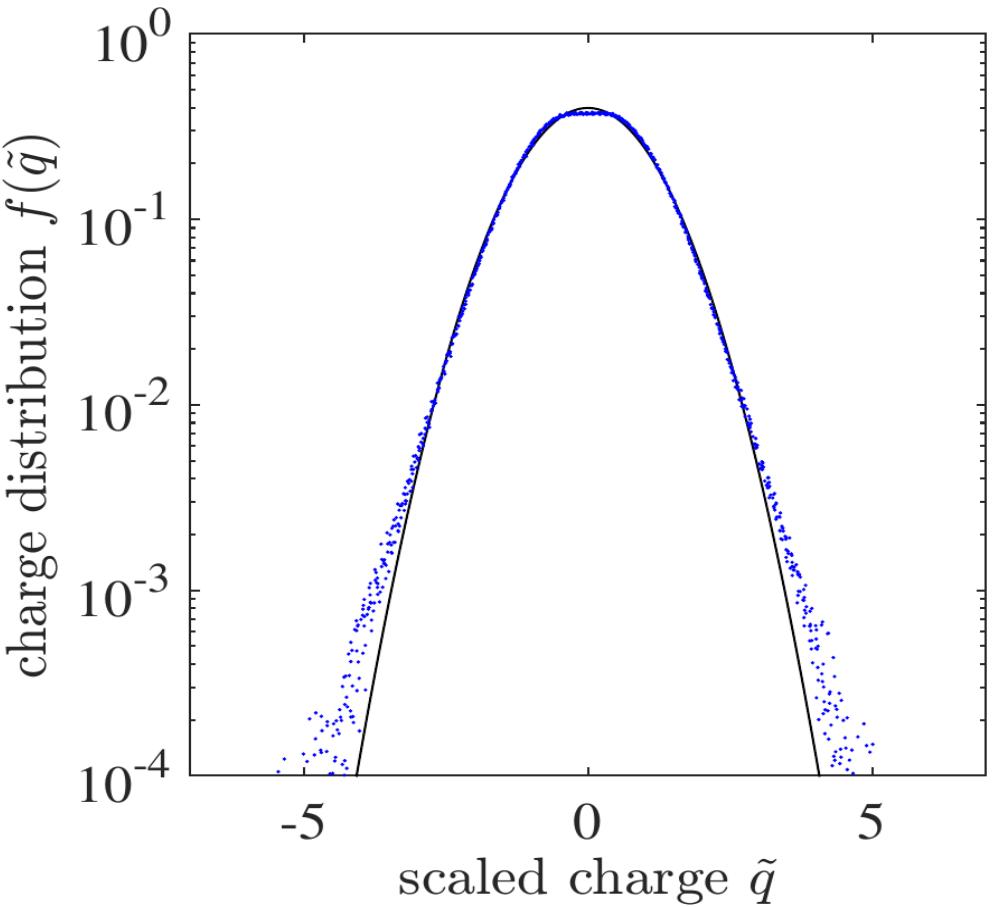}
	\caption{The scaled charge distribution $f(\tilde{q})$ of individual particles obtained from typical MD simulation runs (dots) in the aggregated granular gas. The solid line is a Gaussian fit. Here $\tilde{q}=q/\langle\delta q^2\rangle^{1/2}$.}
	\label{fig_qdist}
\end{figure}	

The attractive or repulsive long-range effects are emulated through an effective differential cross-section for a binary collision, which changes depending upon the sign and magnitude of charges on the particle pair $i,j$ and their relative velocity, according to 
\begin{align}
d\sigma
=  \left(\frac{d\sigma}{d\Omega}\right)d\Omega 
= \frac{d^2}{4}\left( 1 - \frac{2k_e q_i q_j}{d m |\mathbf{v}_{ij}\cdot \mathbf{n}|^2} \right)d\mathbf{n}
\approx \frac{d^2}{4}\left( 1 - \frac{2k_e q_i q_j}{d m v_{ij}^2} \right)d\mathbf{n},
\label{eq_cross_section}
\end{align}
where $\left(\frac{d\sigma}{d\Omega}\right)$ is the differential cross-section per unit solid angle $d\Omega\equiv d\mathbf{n}$. The expression $\frac{d^2}{4}\left( 1 - \frac{2k_e q_i q_j}{d m v_{ij}^2} \right)$ is independent of $\mathbf{n}$, and thus the total cross section is $\sigma = \frac{d^2}{4}\left( 1 - \frac{2k_e q_i q_j}{d m v_{ij}^2} \right)\int_{0}^{\pi} d\phi d\theta \sin \theta = \pi{d^2} \left( 1 - \frac{2k_e q_i q_j}{d m v_{ij}^2} \right) $. For neutral particles, $q_i=q_j=0$, and thus $\sigma = \pi d^2$. For $q_iq_j>0$ (repulsive encounters), $\sigma<\pi d^2$, while for $q_iq_j<0$ (attractive encounters), $\sigma>\pi d^2$. 

Below we explain the solution procedure for the restitutive, as well as aggregative, part of the equation for $\frac{\partial T}{\partial t}$. Similar procedure can then followed for the equations for $\frac{\partial n}{\partial t}$ and $\frac{\partial \langle \delta q^2\rangle}{\partial t}$.

Plugging Eq.~\eqref{eq_delta_psi_T} 
into Eq.~\eqref{eq_psi_splitting} 
and then the resulting equation to Eq.~\eqref{eq_avg_rate_of_change}, 
we find
\begin{align}\nonumber
\frac{3}{2}\frac{\partial T}{\partial t} 
&=
\left( \frac{3}{2}\frac{\partial T}{\partial t} \right)_{res}
+
\left( \frac{3}{2}\frac{\partial T}{\partial t} \right)_{agg}\\ \nonumber
&=
\frac{1}{2}
\int d\mathbf{v}_i
d\mathbf{v}_j
d{q}_i
d{q}_j 
d\sigma
f_i f_j 
|\mathbf{v}_{ij}\cdot \mathbf{n}|
\Theta(-\mathbf{v}_{ij}\cdot \mathbf{n}) \left[ -\frac{1}{2}m(1-\epsilon^2)(\mathbf{v}_{ij}\cdot \mathbf{n})^2\right] \Theta(v_{ij}-\sqrt{b})\\
&+
\frac{1}{2}\int d\mathbf{v}_i
d\mathbf{v}_j
d{q}_i
d{q}_j 
d\sigma
f_i f_j 
|\mathbf{v}_{ij}\cdot \mathbf{n}|
\Theta(-\mathbf{v}_{ij}\cdot \mathbf{n}) \left[ -\frac{1}{2}m(\mathbf{v}_{ij}\cdot \mathbf{n})^2\right] \Theta(-q_iq_j)\Theta(\sqrt{b}-v_{ij}),
\end{align}
which, after using the above Eq.~\eqref{eq_coeff_of_rest} and \eqref{eq_cross_section}, and separating the integrals over $\mathbf{n},\mathbf{v}$ and $q$, reads as
\begin{align}
\frac{3}{2}\frac{\partial T}{\partial t} 
&=
\frac{1}{2}
\int_{q} 
d{q}_i d{q}_j f(q_i) f(q_j)\times {I_\mathbf{v}^{res}}
+
\frac{1}{2}
\int_{q} \Theta(-q_iq_j)
d{q}_i d{q}_j f(q_i) f(q_j) \times {I_\mathbf{v}^{agg}}, 
\label{eq_dTdt_overbraces_start}
\end{align}
where
\begin{align}
{I_\mathbf{v}^{res}}
&=
\int_\mathbf{v} \Theta(v_{ij}-\sqrt{b}) d\mathbf{v}_i d\mathbf{v}_j f(v_i) f(v_j) 
\times
\overbrace{
	\frac{d^2}{4}\left(1-\frac{2k_e q_i q_j}{m d |\mathbf{v}_{ij}|^2}\right)
}^{\frac{d\sigma}{d \Omega}}
\times {I_\mathbf{n}^{res}},
\\
{I_\mathbf{v}^{agg}}
&=
\int_\mathbf{v} \Theta(\sqrt{b}-v_{ij})
d\mathbf{v}_i d\mathbf{v}_j f(v_i) f(v_j) 
\times 	
{
	\frac{d^2}{4}\left(1-\frac{2k_e q_i q_j}{m d |\mathbf{v}_{ij}|^2}\right)
}
\times {I_\mathbf{n}^{agg}}, 
\label{eq_int_I_v_start}
\end{align}		
and
\begin{align}
{I_\mathbf{n}^{res}}
&=	
\int_\mathbf{n} d\mathbf{n} 	|\mathbf{v}_{ij}\cdot \mathbf{n}| \Theta(-\mathbf{v}_{ij}\cdot \mathbf{n}) \left( -m C_\epsilon 	|\mathbf{v}_{ij}\cdot \mathbf{n}|^{11/5} + ... \right), 
\\
{I_\mathbf{n}^{agg}}
&=
{
	\int_\mathbf{n} d\mathbf{n} 	|\mathbf{v}_{ij}\cdot \mathbf{n}| \Theta(-\mathbf{v}_{ij}\cdot \mathbf{n}) \left( -\frac{1}{2}m |\mathbf{v}_{ij}\cdot \mathbf{n}|^{2} \right) 
},
\label{eq_dTdt_overbraces}
\end{align}		
where in the aggregative part, we have set $\epsilon=0$, and $\Theta(-q_iq_j)$ selects only the attractive encounters against low velocities selected by $\Theta(\sqrt{b}-v_{ij})$, the charge-velocity combination which leads to aggregation. Here 
\begin{align}       \sqrt{b}\equiv\sqrt{\frac{2 k_e |q_i q_j|}{md}}.
\end{align} 

\section{Solution for the restitutive part $\left( \frac{3}{2}\frac{\partial T}{\partial t} \right)_{res}$}

The solution for the parts $I_\mathbf{n}^{res}, I_\mathbf{v}^{res}$ are as follows.
\begin{align} \nonumber	
I_\mathbf{n}^{res} &=
\int_\mathbf{n} d\mathbf{n} 	|\mathbf{v}_{ij}\cdot \mathbf{n}| \Theta(-\mathbf{v}_{ij}\cdot \mathbf{n}) \left( -m C_\epsilon 	|\mathbf{v}_{ij}\cdot \mathbf{n}|^{11/5} + ... \right) \\ 
\nonumber	
&=
\int_{0}^{\pi}
\int_{\pi/2}^{\pi} 
d{\phi} d\theta \sin \theta 	
\left( -m C_\epsilon 	|\mathbf{v}_{ij}|^{16/5}  	|\cos \theta|^{16/5} + ... \right) \\ 
&=
-2\pi \frac{5}{21} m C_\epsilon 	|\mathbf{v}_{ij}|^{16/5} + ...	
\label{eq_I_n}
\end{align}
Using Eq.~\eqref{eq_I_n} and \eqref{eq_distributions} from the above text, $I_\mathbf{v}^{res}$ reads as
\begin{align} \nonumber	
I_\mathbf{v}^{res} =
\int_\mathbf{v} \Theta(v_{ij}-\sqrt{b})
d\mathbf{v}_i d\mathbf{v}_j &
\left[n\left(\frac{m}{2\pi T}\right)^{3/2} e^{-(m/2T)v_i^2}\right] 
\left[n\left(\frac{m}{2\pi T}\right)^{3/2} e^{-(m/2T)v_j^2}\right]\\
&\times
\left[
-2\pi \frac{5}{21} m C_\epsilon 	|\mathbf{v}_{ij}|^{16/5}\right]
\frac{d^2}{4}\left(1-\frac{2k_e q_i q_j}{m d |\mathbf{v}_{ij}|^2}\right).
\end{align}
To perform the integration over the relative velocity $\mathbf{v}_{ij}$, the following transformations are made: (i) $\mathbf{w}_{ij}=\frac{2T}{m}(\mathbf{v}_{i}-\mathbf{v}_{j})$, and (ii) $\mathbf{w}_{c}=\frac{2T}{m}(\mathbf{v}_{i}+\mathbf{v}_{j})$, which also results in $d\mathbf{v}_{i}d\mathbf{v}_{j}=-\frac{1}{8}\left(\frac{m}{4T}\right)^{-3}d\mathbf{w}_{ij}d\mathbf{w}_{c}$. Incorporating these transformations, we obtain
\begin{align} \nonumber	
I_\mathbf{v}^{res}&=		
\left[ - \frac{n^2 d^2}{4} \frac{1}{8} \left(\frac{m}{4T}\right)^{-3} \left(\frac{m}{2\pi T}\right)^{3} \left(\frac{4T}{m}\right)^{16/10} \right]\\\nonumber
&\times
\int_{\mathbf{w}_{ij}} 	
d\mathbf{w}_{ij} 
\left[
\int_{\mathbf{w}_c} 
d\mathbf{w}_c e^{-w_c^2}
\right] 		
e^{-w_{ij}^2}
\left[
-2\pi \frac{5}{21} m C_\epsilon 	{w}_{ij}^{16/5}\right]
\left(1-\frac{k_e q_i q_j}{2 T d w_{ij}^2}\right)\\\nonumber
&=
-2 T^{8/5} \left[\frac{2^{1/5} n^2 d^2 }{\pi^3 m^{8/5}} \right]
\underbrace{
	\left[
	4 \pi \int_{0}^{\infty}
	d{w}_c w_c^2 e^{-w_c^2}
	\right]
}_{I_{w_c}} \\
&\times
\underbrace{
	4 \pi \int_{\sqrt{\frac{k_e |q_i q_j|}{2Td}}}^{\infty} 	
	d{w}_{ij} w_{ij}^2 		
	e^{-w_{ij}^2}
	\left[
	-2\pi \frac{5}{21} m C_\epsilon 	{w}_{ij}^{16/5}\right] 
	\left(1-\frac{k_e q_i q_j}{2 T d w_{ij}^2}\right)
}_{I_{w_{ij}}}.
\label{eq_int_I_v}
\end{align}		
Notice the lower limit on relative velocities,
\begin{align}      \sqrt{b}\equiv\sqrt{\frac{k_e |q_i q_j|}{2Td}},
\end{align} 	%
is also altered due to the transformation $\mathbf{v}_{ij}\rightarrow\mathbf{w}_{ij}$. The integral $I_{w_c}$ gives
\begin{align}
I_{w_{c}}&=\pi^{3/2},
\end{align}
while the integral $I_{w_{ij}}$ is solved as
\begin{align}\nonumber
I_{w_{ij}} &= 
-\frac{40 \pi^2 }{21} m C_\epsilon 
\int_{\sqrt{\frac{k_e |q_i q_j|}{2Td}}}^{\infty} 
d{w}_{ij} 	
{w}_{ij}^{26/5}	
e^{-w_{ij}^2}		
\left(1-\frac{k_e q_i q_j}{2 T d w_{ij}^2}\right)\\\nonumber
&= 
-\frac{40 \pi^2 }{21} m C_\epsilon 
\int_{\sqrt{b}}^{\infty} 
d{w}_{ij} 	
{w}_{ij}^{26/5}	
e^{-w_{ij}^2}		
\left(1-\frac{a}{w_{ij}^2}\right)\\\nonumber
&=
-\frac{40 \pi^2 }{21} m C_\epsilon
\frac{1}{2}\left[
\Gamma\left(\frac{31}{10},b\right)
-a\Gamma\left(\frac{21}{10},b\right)
\right]\\\nonumber		
&\approx
-\frac{40 \pi^2 }{21} m C_\epsilon
\frac{1}{2}\left[
\Gamma\left(3,b\right)
-a\Gamma\left(2,b\right)
\right]\\	
&=
-\frac{40 \pi^2 }{21} m C_\epsilon
\frac{1}{2}\left[
2e^{-b}(1+b+b^2)
-a e^{-b}(1+b+b^2)
\right],
\end{align}
where $a={\frac{k_e q_i q_j}{2Td}}$ and ${b}={\frac{k_e |q_i q_j|}{2Td}}$. Putting $I_{w_c},I_{w_{ij}}$ in $I_\mathbf{v}^{res}$, and finally $I_\mathbf{v}^{res}$ in the restitutive part of Eq.~\eqref{eq_dTdt_overbraces_start}, and integrating over $q_i,q_j$, we obtain
\begin{align}\nonumber
\left(\frac{3}{2}\frac{\partial T}{\partial t} \right)_{res}
&=
\frac{1}{2}
\int_{-\infty}^{\infty}
\int_{-\infty}^{\infty} 
d{q}_i d{q}_j 
\frac{1}{\sqrt{2\pi \langle \delta q^2 \rangle}} e^{-q_i^2 / (2 \langle \delta q^2 \rangle)}
\frac{1}{\sqrt{2\pi \langle \delta q^2 \rangle}} e^{-q_j^2 / (2 \langle \delta q^2 \rangle)}\\\nonumber
&\times
(-2) T^{8/5} \left[\frac{2^{1/5} n^2 d^2 }{\pi^3 m^{8/5}} \right]				
\pi^{3/2}
\frac{(-40 \pi^2) }{21} m C_\epsilon
\frac{1}{2}\left[
2e^{-b}(1+b+b^2)
-a e^{-b}(1+b+b^2)
\right]\\ 
&=
-T^{8/5} \left[\frac{C^T_{res}}{l^5}\right] \left[a^T_1 + a^T_2\tan^{-1}\frac{\mathcal{B}}{l}\right].		
\end{align}
Notice that if $\mathcal{B}=0$, we recover the classical Haff's law for viscoelastic granular gas. 

\section{Solution for the aggregative part $\left( \frac{3}{2}\frac{\partial T}{\partial t} \right)_{agg}$}	

The solution for the parts $I_\mathbf{n}^{agg}, I_\mathbf{v}^{agg}$ are as follows.
\begin{align} \nonumber	
I_\mathbf{n}^{agg} &=
\int_\mathbf{n} d\mathbf{n} 	|\mathbf{v}_{ij}\cdot \mathbf{n}| \Theta(-\mathbf{v}_{ij}\cdot \mathbf{n}) \left( -\frac{1}{2} m |\mathbf{v}_{ij}\cdot \mathbf{n}|^{2}\right) \\ 
\nonumber	
&=
\int_{0}^{\pi}
\int_{\pi/2}^{\pi} 
d{\phi} d\theta \sin \theta 
|\mathbf{v}_{ij}| |\cos \theta|
\left( -\frac{1}{2} m |\mathbf{v}_{ij}|^{2}  	|\cos \theta|^{2}\right) \\ 
&=
\frac{\pi}{4} m 	|\mathbf{v}_{ij}|^{3}.
\label{eq_I_n_agg}
\end{align}
Using Eq.~\eqref{eq_I_n_agg} and \eqref{eq_distributions} from the above text, and again using the variable transformations (i) $\mathbf{w}_{ij}=\frac{2T}{m}(\mathbf{v}_{i}-\mathbf{v}_{j})$, (ii) $\mathbf{w}_{c}=\frac{2T}{m}(\mathbf{v}_{i}+\mathbf{v}_{j})$, (iii) $d\mathbf{v}_{i}d\mathbf{v}_{j}=-\frac{1}{8}\left(\frac{m}{4T}\right)^{-3}d\mathbf{w}_{ij}d\mathbf{w}_{c}$, the integral $I_\mathbf{v}^{agg}$ in~\eqref{eq_int_I_v_start} reduces to
\begin{align}
I_\mathbf{v}^{agg}		
&=
- T^{3} \left[\frac{2\sqrt{2} \pi^2 d^2}{m^2} \right]
\underbrace{
	\int_{0}^{\sqrt{\frac{k_e |q_i q_j|}{2Td}}} 	
	d{w}_{ij} w_{ij}^2 	w_{ij}^3	
	e^{-w_{ij}^2}
	\left(1-\frac{k_e q_i q_j}{2 T d w_{ij}^2}\right)
}_{I_{w_{ij}}}.
\label{eq_int_I_v_agg}
\end{align}	
Here notice that now the relative velocity limits are from $v_{ij}=0$ to $\sqrt{b}$, the condition for aggregation selected by $\Theta(\sqrt{b}-v_{ij})$. Finally putting~\eqref{eq_int_I_v_agg} into aggregative part of~\eqref{eq_dTdt_overbraces_start} and integrating over $q_i,q_j$, we obtain
\begin{align}
\left(\frac{3}{2}\frac{\partial T}{\partial t} \right)_{agg}
&=
n^2T^{3/2}
\left[\frac{C^T_{agg}}{l^5}\right]
\left[
a^T_3 + a^T_4 \cot^{-1}\frac{\mathcal{B}}{l}
+a^T_5 + a^T_6 \tan^{-1}\frac{\mathcal{B}}{l} \right].
\end{align}
Notice that after integrating from $v_{ij}=0$ to $\sqrt{b}$, the integration over $q_i,q_j$ is to be broken into the sum of two parts,
one over $q_i\in (-\infty, 0]$, $q_j \in [0,+\infty)$, plus  a second integral over $q_i\in [0, +\infty)$, $q_j \in(-\infty, 0]$, to satisfy the aggregative condition set by $\Theta(-q_iq_j)\Theta(\sqrt{b}-v_{ij})$.

A similar procedure is followed for 		$\left(\frac{\partial \langle\delta q^2\rangle}{\partial t} \right)_{res}$, $\left(\frac{\partial \langle\delta q^2\rangle}{\partial t} \right)_{agg}$ and
$\left(\frac{\partial n}{\partial t} \right)_{agg}$ using corresponding $\Delta[\psi_i]$. Finally, the key constraint to be noted is that the solutions of the integrals in the rate equation for $T$ are real valued for $\mathcal{B}\leq 4$, while in equations for $\langle \delta q^2 \rangle$ and $n$, they are real valued for $\mathcal{B}\leq 1$. The MD simulations confirm that these constraints put a physical limit during the aggregation phase.

\begin{table}[t!]
	\centering
	\renewcommand{\arraystretch}{0.7}
	\begin{tabular}{>{\centering}m{0.3\textwidth}    >{\centering\arraybackslash}m{0.7\textwidth}}
		\hline
		Coefficient & Expression\\ 
		\hline
		$\mathcal{B}$         & ${k_e \langle \delta q ^2 \rangle}/{(T d)}$ \\
		$l$         & $\sqrt{4-\mathcal{B}^2}$\\ 
		$l_1$       & $\sqrt{1-\mathcal{B}^2}$\\ 			
		$C^T_{res}$	& $80\;{2^{1/5}\; d^2 C_\epsilon m}/{(21                     \sqrt{\pi}m^{8/5})}$\\
		$C^T_{agg}$	& $ d^2/(8\sqrt{\pi m})$\\
		$C^q_{res}$	& $1.4080\;  d^2  C_{\Delta q}^2 (2+\eta)(4/m)^{\eta+1/2}/\pi$\\
		$C^q_{agg}$	& $4{ d^2}/\sqrt{\pi m}$\\
		$C^n_{agg}$	& ${ d^2}/(2\sqrt{\pi m})$\\\\
		
		$a^T_1$     & $\mathcal{B}(8-5\mathcal{B}^2)l + \pi(16-10\mathcal{B}^2+3\mathcal{B}^4)$\\ 
		$a^T_2$     & $-32+20\mathcal{B}^2-6\mathcal{B}^4$\\
		$a^T_3$     & $l(2\pi (\mathcal{B}^2-4)^2+\mathcal{B}(-32+28\mathcal{B}^2+\mathcal{B}^4))$\\
		$a^T_4$     & $-4(32-20\mathcal{B}^2+9\mathcal{B}^4)$\\
		$a^T_5$     & $l\mathcal{B}(-32+28\mathcal{B}^2+\mathcal{B}^4)$\\
		& $+2\pi(\mathcal{B}^2(20-8l)+\mathcal{B}^4(-9+l)+16(-2+l))$\\
		$a^T_6$     & $4(32-20\mathcal{B}^2+9\mathcal{B}^4)$\\  \\
		
		$a^q_1$     & $2\mathcal{B}l(\mathcal{B}^2-1)-\pi(4-2\mathcal{B}^2+\mathcal{B}^4)$\\
		$a^q_2$     & $2(4-2\mathcal{B}^2+\mathcal{B}^4)$\\
		$a^q_3$     & $16l \sqrt{\pi} l_1 (4-5\mathcal{B}^2+\mathcal{B}^4)^2$\\
		$a^q_4$     & $-\sqrt{\pi}\mathcal{B}l(\pi(\mathcal{B}^2-4)(1+2\mathcal{B}^2)$\\
		& $+2\mathcal{B}l_1(-8-54\mathcal{B}^2+33\mathcal{B}^4+2\mathcal{B}^6) )$\\
		$a^q_5$     & $2\sqrt{\pi}\mathcal{B}l^5(1+2\mathcal{B}^2)$\\
		$a^q_6$     & $16\sqrt{\pi}\mathcal{B} l_1^5 (4+5\mathcal{B}^2)$\\
		$a^q_7$     & $8l \sqrt{\pi} l_1 (4-5\mathcal{B}^2+\mathcal{B}^4)^2$\\
		$a^q_8$     & $-\sqrt{\pi}\mathcal{B}l^5(1+2\mathcal{B}^2)$\\
		$a^q_9$     & $\sqrt{\pi}\mathcal{B}l_1 \mathcal{B} l (8+54\mathcal{B}^2-33\mathcal{B}^4-2\mathcal{B}^6)$\\
		$a^q_{10}$     & $8\sqrt{\pi}\mathcal{B}l_1 (\mathcal{B}^2-1)^2 (4+5\mathcal{B}^2)$\\  \\
		
		$a^n_1$     & $-\mathcal{B}^2l^3$\\
		$a^n_2$     & $l_1l \mathcal{B}(-4+7\mathcal{B}^2)+l_1\pi (\mathcal{B}^2-1)(8-4l+\mathcal{B}^2 (-6+l))$\\
		$a^n_3$     & $4l_1(4-7\mathcal{B}^2+3\mathcal{B}^4)$\\
		$a^n_4$     & $-\mathcal{B}^2l^3$\\
		$a^n_5$     & $l_1l(\mathcal{B}(-4+7\mathcal{B}^2)+\pi(4-5\mathcal{B}^2+\mathcal{B}^4))$\\
		$a^n_6$     & $-4l_1(4-7\mathcal{B}^2+3\mathcal{B}^4)$\\ 		 
		\hline
		
	\end{tabular}
	\caption{Expressions of the coefficients in Eq.~(5)-(7) in the main text. Here $m$ and $d$ are the mass and size of the aggregates. The material constant $C_\epsilon$ is from Eq.~\eqref{eq_coeff_of_rest} and influence the viscoelastic properties of the particles, while $C_{\Delta q}$ and $\eta$ are from Eq.~\eqref{eq_dq} and influence the charge buildup. Other notations are as described in the main text.}
	\label{table_coeffs}
\end{table}	

\section{Granular MD simulations} 
The equation of motion of the form
\begin{align}\label{eq_motion_nondim}\nonumber
\frac{d\mathbf{v}_i}{dt}  
=&  
\sum_j\left[ \Theta(\xi_{ij})\left(\mathcal{E} \xi_{ij}^{\frac{3}{2}}- \mathcal{D}\xi_{ij}^{\frac{1}{2}} \dot{\xi}_{ij}     \right)\mathbf{n}_{ji} \right]\\&
+ \mathcal{K}  
\sum_{k}
\sum_{\mathbf{b}} 
{{}^\prime}
\frac{q_i q_k}{|\mathbf{r}_{ki}+\mathbf{b}L|^{3}} (\mathbf{r}_{ki}+\mathbf{b}L),
\end{align}
is solved for each particle with a setup of periodic boundary conditions in a cubic box of size $L^3=70d_0\times70d_0\times70d_0$, where $d_0$ is the monomer diameter [Table~\ref{table_params},~\ref{table_sim_params}]. Here $ \mathbf{b}$ is a vector of integers representing the periodic replicas of the system in each Cartesian direction. The symbol $'$ indicates that $k\neq i$ if $\mathbf{b}=0$ to avoid Coulomb interaction of particles with themselves.
The non-dimensional numbers in the above equation are $\mathcal{E}=\frac{\alpha l_\text{ref}^{3/2}t_\text{ref}}{m_\text{ref} v_\text{ref}}$, $\mathcal{D}=\frac{\beta l_\text{ref}^{1/2}t_\text{ref}}{m_\text{ref}}$ and $\mathcal{K}=\frac{k_e q_\text{ref}^{2}t_\text{ref}}{m_\text{ref} v_\text{ref} l_\text{ref}^2}$, with $\alpha$ and $\beta$ being viscoelatic material constants. From practical problems, we select the reference length $l_\textrm{ref}\equiv d_0$, time reference $t_\textrm{ref}$, velocity reference $v_\textrm{ref}$, and charge reference $q_\textrm{ref}$ such that the elastic force strength $\mathcal{E}\approx 278$, dissipative force strength $\mathcal{E}/10$, and Coulomb force strength is varied across $\mathcal{K}=0.4-5.0$ [Table~\ref{table_params},~\ref{table_sim_params}]. The effect of dissipation compared to elastic forces is extensively studied for neutral systems~\cite{brilliantov2010kinetic}. The variation of Coulomb strength compared to dissipation and elastic forces we have repoted in~\cite{singh2018early}. The above order of magnitudes of $\mathcal{E}$, $\mathcal{D}$, and $\mathcal{K}$ also helps to attain an early clustering in non-dimensional time units in a finite size ($N\sim50000$) neutral granular gas system [see \cite{singh2018early} for more details]. Also $\xi_{ij}\equiv d_0-r_{ij},\; \dot{\xi}_{ij}\equiv \frac{d\xi}{dt}$, $\mathbf{n}_{ji}$ is the unit vector pointing from center of in-contact neighbor $j$ towards the center of particle $i$, while  $\mathbf{r}_{ki}$ is the distance vector pointing from particle $k$ towards the center of particle $i$. 

The equation of charge on particle $i$ may be written as
\begin{align}
\frac{dq_i}{dt} =&  
\sum_j\left[ \Theta(\xi_{ij}) I_{ji}  \right],
\end{align}
with $I_{ji}$ being the charge-exchange currents from colliding neighbors $j$ during the course of collision. For any contact neighbor $j$, we approximate its integrated value over the time step $\tau$ by Eq.~\eqref{eq_dq}, {\it i.e.}
\begin{align}
(q_i'-q_i) = \int dq_i = \int_{0}^{\tau} I_{ji}  dt
\approx
C_{\Delta q} |\mathbf{v}_{ij}\cdot \mathbf{n}|^\eta \frac{q_i-q_0}{|q_i-q_0|}.
\end{align} 
After charge-exchange, the long-range Coulomb forces for a setup with periodic boundary conditions in Eq.~\eqref{eq_motion_nondim} is challenging and conditionally convergent as it depends on the order of summation. We employ the Ewald summation that converges rapidly, and has a computational complexity $\mathcal{O}(N^{3/2})$. The algorithm is highly parallelized and optimized on graphics processing unit (GPU). In our simulations, the total computing time to reach non-dimensional simulation time $\sim 10^3$ for a typical simulation with monomers $N \sim 10^5$, including the long-range electrostatic forces, is of the order of weeks. See \cite{singh2018early} for more details.

\section{Comparison of individual $T$, $\langle \delta q^2 \rangle$, and $n$ profiles, and emergence of convective flow using Mach number}
\begin{figure}
	\includegraphics[width=1.0\linewidth]{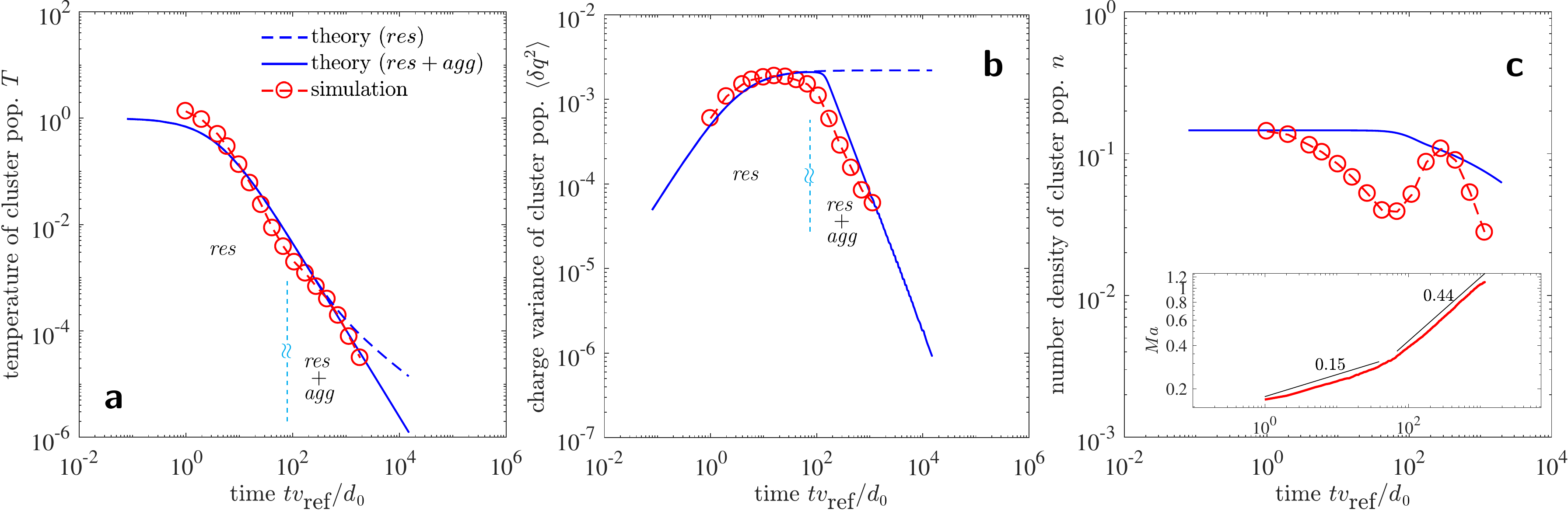}
	\caption{Evolution of ({a}) temperature of cluster population $T$, 
		({b}) charge variance of cluster population $\langle\delta q^2\rangle$, and ({c}) number density of cluster population $n$. The granular temperature, charge variance and average size of the cluster population during aggregation evolve in such a manner that their non-dimensional combination $\mathcal{B}(t)=k_e\langle \delta q^2\rangle/(Td)  \le 1$ (main text). The number density evolution, however, is highly dynamic and exhibits a non-monotonic behavior due to emergence of macroscopic flow, quantified by the Mach number $Ma$ ({c}) inset). The short-hands $res$ and $agg$ denote restitution and aggregation respectively.}
	\label{fig_T_qq_n_supp}
\end{figure}
In Fig.~\ref{fig_T_qq_n_supp}, we decompose the theoretical comparison of $\mathcal{B}$, presented in the main text, into individual comparisons of $T$, $\langle \delta q^2 \rangle$, and $n$ profiles for a typical simulation run. The difference between the kinetic theory with and without aggregation is also emphasized. 

It is noticeable that the granular temperature of the aggregates closely follows Haff's law, and is confirmed by theory, notwithstanding the presence of long-range effects and intricate aggregation and annihilation events. If only the restitutive terms of the hydrodynamic equations are considered (dashed line), the theory predicts that $T$ drops at a slower rate at long times. Furthermore, the charge variance in this case saturates. The number density in the absence of aggregation is, of course, invariant. If the aggregation dynamics is augmented, the simulation results are closely predicted by the theory.

In the MD simulations, the decay of $\langle \delta q^2 \rangle$ during the aggregation phase closely agrees with the theory, even though we observe that the charge exchange in the simulations leads to a symmetric but non-Gaussian charge distribution among monomers during the initial restitution phase [Fig.~1].  

The number density evolution, however, is highly dynamic and exhibits a non-monotonic behavior due to fragmentation event caused by the emergence of macroscopic flow. The theory predicts the decay of cluster density only in an average sense [Fig. \ref{fig_T_qq_n_supp}(c) inset]. To quantify the emergence of the macroscopic flow, we calculate the local Mach number
\begin{align}
Ma={\frac{\langle\mathbf{V}^2\rangle^{1/2}}{v_T}}\,,
\end{align}
where $\mathbf{V}$ is the macroscopic velocity, and $v_T$ the thermal velocity.
The time evolution of ${Ma}$ is shown in Fig.~\ref{fig_T_qq_n_supp}(c) (inset), which indicates that ${Ma}$ grows at a higher rate when the number density evolution becomes non-monotonic ($tv_\textrm{ref}/d_0 \approx 10^2$), indicating an intricate aggregation and fragmentation dynamics, and the generation of macroscopic flow.

\section{REFERENCE SCALES AND Laboratory relevance of present results}
\begin{figure}\centering
	\includegraphics[width=0.38\textwidth]{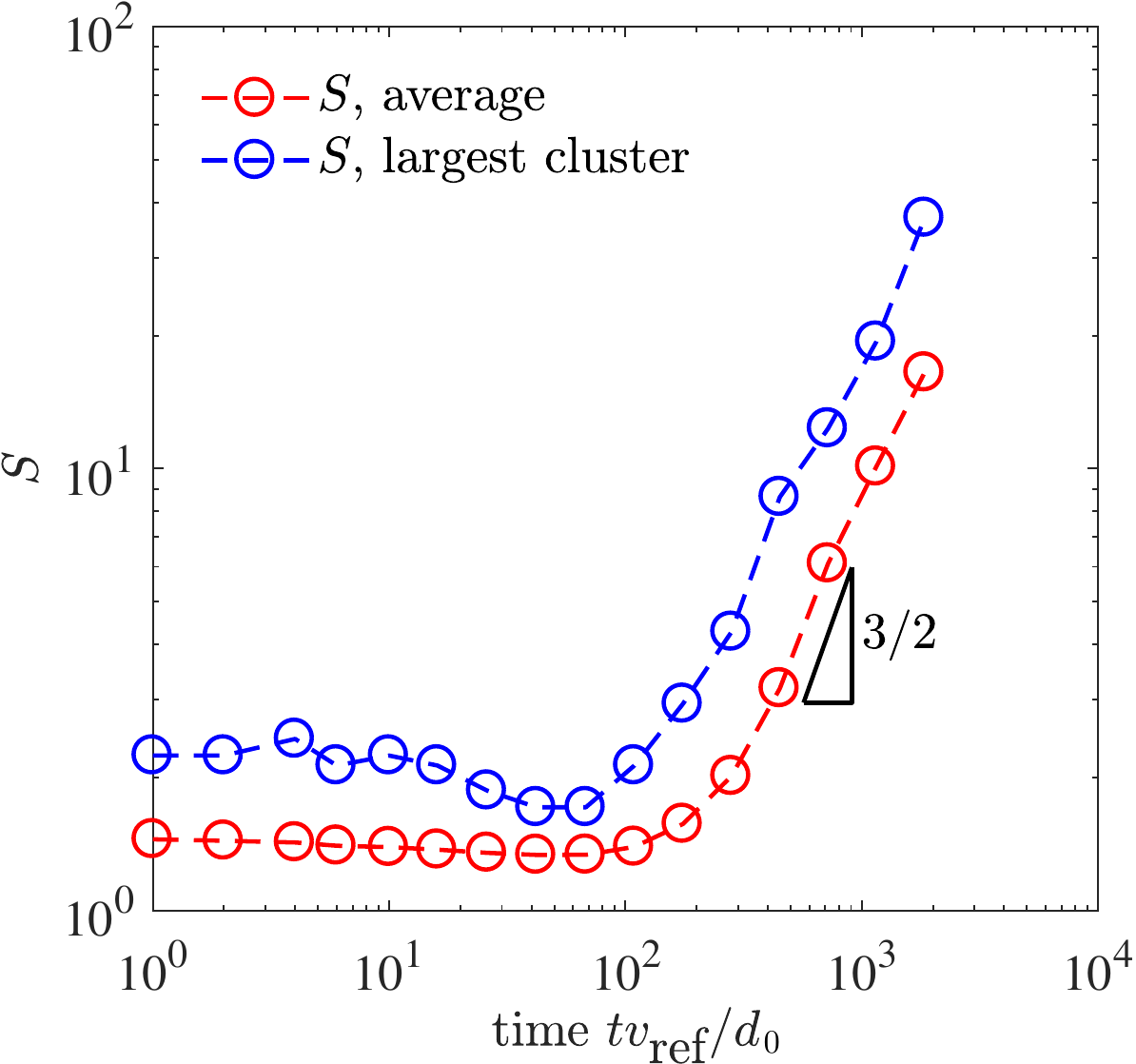}
	\caption{Growth of the average cluster size, and the size of the largest cluster.}
	\label{fig_rg}
\end{figure} 
Typically non-Brownian growth in planetary dust becomes dominant for monomer sizes near or above several $\si{\mu m}$ \cite{zsom2010outcome} and the growth barrier problem \cite{spahn2015granular} starts to arise near $d\sim 10^{-3}\;\si{m}$. The mass of silica particles in this range of sizes is $m\sim 10^{-4}-10^{-6}\; \si{kg}$. If the particles are initially agitated with velocities $v\sim 1.0 \; \si{m \;s^{-1}}$, the time scale reference to convert our simulation time to laboratory time is $d/v_{\mathrm{ref}}\sim 4.78\times 10^{-3}\;\si{s}$. Thus in our results the growth over $10^4$ units of non-dimensional time approximately implies growth over $\sim 10\;\si{s}$. The average size of aggregates in the growth period grows approximately by one order of magnitude ({\it e.g.} the growth of the largest cluster is from $\approx 2 \; \si{mm}$ to $\approx 7 \; \si{cm}$ in $\approx 10\;\si{s}$ time for particles of such size and mass, and for initial monomer filling fraction of $\phi=0.076$) [Fig.~\ref{fig_rg}].

\begin{table}[t!]
	\centering
	\renewcommand{\arraystretch}{0.8}
	\begin{tabular}{>{\centering}m{0.3\textwidth}    >{\centering\arraybackslash}m{0.4\textwidth}
			>{\centering\arraybackslash}m{0.3\textwidth}}
		\hline
		Parameter & Expression &Value\\ 
		\hline
		
		$N_\text{mon}$ & No. of monomers  & $50016$\\ 
		$L$            & System length    & $70$\\
		$\phi$         & Monomer filling fraction & 0.013, 0.030, 0.076, 0.137\\ 
		
		$\mathcal{E}$  & $\frac{\alpha l_\text{ref}^{3/2}t_\text{ref}}{m_\text{ref} v_\text{ref}}$ & 278\\
		$\mathcal{D}$ & $\frac{\beta l_\text{ref}^{1/2}t_\text{ref}}{m_\text{ref}}$ & 27.8\\
		$\mathcal{K}$ & $\frac{k_e q_\text{ref}^{2}t_\text{ref}}{m_\text{ref} v_\text{ref} l_\text{ref}^2}$ & 0.4, 1.0, 5.0\\ [2ex]
		\hline
	\end{tabular}
	\caption{Simulation parameters.}
	\label{table_params}
\end{table}	
\begin{table}[t!]
	\centering
	\renewcommand{\arraystretch}{0.8}
	\begin{tabular}{>{\centering}m{0.3\textwidth}    >{\centering\arraybackslash}m{0.4\textwidth}
			>{\centering\arraybackslash}m{0.3\textwidth}}
		\hline
		Reference scale & Description &Value\\ 
		\hline
		$m_\text{ref}$ & Particle mass reference & $1.52\times 10^{-4}$ \si{kg}\\
		$l_\text{ref}=d_0$ & Length reference=monomer diameter & $4.78\times 10^{-3}$ \si{m}\\
		$v_\text{ref}$ & Velocity reference & $1.0$ \si{m s^{-1}} \\	
		$t_\text{ref}=d_0/v_\text{ref}$ & Time reference reference & $4.78\times 10^{-3}$ \si{s}\\
		$T_\text{ref}=\frac{1}{3} m_\text{ref} v_\text{ref}^2$ & Temperature reference & $0.51\times 10^{-4}$ $\textrm{kg} \;\textrm{m}^2\; \textrm{s}^{-2}$\\
		$k_e$ & Coulomb's constant & $8.98\times10^9$ $\textrm{N} \;\textrm{m}^2\; \textrm{C}^{-2}$\\
		$\alpha$ & Elastic constant of particles & $2.67\times 10^4$ $\textrm{kg}\;\textrm{m}^{-1/2}\textrm{s}^{-2}$\\
		$\beta$  & Viscous constant of particles & $1.28\times 10^1$ $\textrm{kg}\;\textrm{m}^{-1/2}\textrm{s}^{-1}$\\
		$q_\text{ref}$ & Charge reference & $2.0\times 10^{-8}$ $\textrm{C}$ - $5.6\times 10^{-9}$ $\textrm{C}$\\
		\hline
	\end{tabular}
	\caption{Reference values in SI units for non-dimensionalization.}
	\label{table_sim_params}
\end{table}	

\section{Codes and media}
The following computer codes and media are made public:
\begin{enumerate}
	\item \texttt{MATLAB} program to solve Eq.~(5)-(7).
	
	\item \texttt{MATHEMATICA} framework to solve the restitutive and aggregative parts of the kinetic integrals.
	
	\item The cluster analysis code in \texttt{MATLAB} to obtain the fractal dimension, cluster size distribution, average cluster size, and other statistical quantities, is made available at \href{https://gitlab.com/cphyme/cluster-analysis}{https://gitlab.com/cphyme/cluster-analysis}.
	
	\item Movie of the evolving and aggregating structures in the charged granular gas.
\end{enumerate}

\section{Data availability}
All data are available from the authors.	
\end{widetext}

\section{Acknowledgements}
We gratefully acknowledge highly constructive comments by J\"urgen Blum, Eberhard Bodenschatz, Stephan Herminghaus, and Reiner Kree. We thank the Max Planck Society for funding.

%
%
%

\begin{thebibliography}{35}%
\makeatletter
\providecommand \@ifxundefined [1]{%
 \@ifx{#1\undefined}
}%
\providecommand \@ifnum [1]{%
 \ifnum #1\expandafter \@firstoftwo
 \else \expandafter \@secondoftwo
 \fi
}%
\providecommand \@ifx [1]{%
 \ifx #1\expandafter \@firstoftwo
 \else \expandafter \@secondoftwo
 \fi
}%
\providecommand \natexlab [1]{#1}%
\providecommand \enquote  [1]{``#1''}%
\providecommand \bibnamefont  [1]{#1}%
\providecommand \bibfnamefont [1]{#1}%
\providecommand \citenamefont [1]{#1}%
\providecommand \href@noop [0]{\@secondoftwo}%
\providecommand \href [0]{\begingroup \@sanitize@url \@href}%
\providecommand \@href[1]{\@@startlink{#1}\@@href}%
\providecommand \@@href[1]{\endgroup#1\@@endlink}%
\providecommand \@sanitize@url [0]{\catcode `\\12\catcode `\$12\catcode
  `\&12\catcode `\#12\catcode `\^12\catcode `\_12\catcode `\%12\relax}%
\providecommand \@@startlink[1]{}%
\providecommand \@@endlink[0]{}%
\providecommand \url  [0]{\begingroup\@sanitize@url \@url }%
\providecommand \@url [1]{\endgroup\@href {#1}{\urlprefix }}%
\providecommand \urlprefix  [0]{URL }%
\providecommand \Eprint [0]{\href }%
\providecommand \doibase [0]{http://dx.doi.org/}%
\providecommand \selectlanguage [0]{\@gobble}%
\providecommand \bibinfo  [0]{\@secondoftwo}%
\providecommand \bibfield  [0]{\@secondoftwo}%
\providecommand \translation [1]{[#1]}%
\providecommand \BibitemOpen [0]{}%
\providecommand \bibitemStop [0]{}%
\providecommand \bibitemNoStop [0]{.\EOS\space}%
\providecommand \EOS [0]{\spacefactor3000\relax}%
\providecommand \BibitemShut  [1]{\csname bibitem#1\endcsname}%
\let\auto@bib@innerbib\@empty
\bibitem [{\citenamefont {Castellanos}(2005)}]{castellanos2005relationship}%
  \BibitemOpen
  \bibfield  {author} {\bibinfo {author} {\bibfnamefont {A}~\bibnamefont
  {Castellanos}},\ }\bibfield  {title} {\enquote {\bibinfo {title} {The
  relationship between attractive interparticle forces and bulk behaviour in
  dry and uncharged fine powders},}\ }\href@noop {} {\bibfield  {journal}
  {\bibinfo  {journal} {Advances in Physics}\ }\textbf {\bibinfo {volume}
  {54}},\ \bibinfo {pages} {263--376} (\bibinfo {year} {2005})}\BibitemShut
  {NoStop}%
\bibitem [{\citenamefont {Schwager}\ \emph {et~al.}(2008)\citenamefont
  {Schwager}, \citenamefont {Wolf},\ and\ \citenamefont
  {P{\"o}schel}}]{schwager2008fractal}%
  \BibitemOpen
  \bibfield  {author} {\bibinfo {author} {\bibfnamefont {Thomas}\ \bibnamefont
  {Schwager}}, \bibinfo {author} {\bibfnamefont {Dietrich~E}\ \bibnamefont
  {Wolf}}, \ and\ \bibinfo {author} {\bibfnamefont {Thorsten}\ \bibnamefont
  {P{\"o}schel}},\ }\bibfield  {title} {\enquote {\bibinfo {title} {Fractal
  substructure of a nanopowder},}\ }\href@noop {} {\bibfield  {journal}
  {\bibinfo  {journal} {Physical Review Letters}\ }\textbf {\bibinfo {volume}
  {100}},\ \bibinfo {pages} {218002} (\bibinfo {year} {2008})}\BibitemShut
  {NoStop}%
\bibitem [{\citenamefont {Wesson}(1973)}]{wesson1973accretion}%
  \BibitemOpen
  \bibfield  {author} {\bibinfo {author} {\bibfnamefont {Paul~S}\ \bibnamefont
  {Wesson}},\ }\bibfield  {title} {\enquote {\bibinfo {title} {Accretion and
  electrostatic interaction of interstellar dust grains; interstellar grit},}\
  }\href@noop {} {\bibfield  {journal} {\bibinfo  {journal} {Astrophysics and
  Space Science}\ }\textbf {\bibinfo {volume} {23}},\ \bibinfo {pages}
  {227--255} (\bibinfo {year} {1973})}\BibitemShut {NoStop}%
\bibitem [{\citenamefont {Harper}\ \emph {et~al.}(2017)\citenamefont {Harper},
  \citenamefont {McDonald}, \citenamefont {Dufek}, \citenamefont {Malaska},
  \citenamefont {Burr}, \citenamefont {Hayes}, \citenamefont {McAdams},\ and\
  \citenamefont {Wray}}]{harper2017electrification}%
  \BibitemOpen
  \bibfield  {author} {\bibinfo {author} {\bibfnamefont {J~S~M{\'e}ndez}\
  \bibnamefont {Harper}}, \bibinfo {author} {\bibfnamefont {G~D}\ \bibnamefont
  {McDonald}}, \bibinfo {author} {\bibfnamefont {J}~\bibnamefont {Dufek}},
  \bibinfo {author} {\bibfnamefont {M~J}\ \bibnamefont {Malaska}}, \bibinfo
  {author} {\bibfnamefont {D~M}\ \bibnamefont {Burr}}, \bibinfo {author}
  {\bibfnamefont {A~G}\ \bibnamefont {Hayes}}, \bibinfo {author} {\bibfnamefont
  {J}~\bibnamefont {McAdams}}, \ and\ \bibinfo {author} {\bibfnamefont {J~J}\
  \bibnamefont {Wray}},\ }\bibfield  {title} {\enquote {\bibinfo {title}
  {Electrification of sand on titan and its influence on sediment transport},}\
  }\href@noop {} {\bibfield  {journal} {\bibinfo  {journal} {Nature
  Geoscience}\ }\textbf {\bibinfo {volume} {10}},\ \bibinfo {pages} {260}
  (\bibinfo {year} {2017})}\BibitemShut {NoStop}%
\bibitem [{\citenamefont {Brilliantov}\ \emph {et~al.}(2015)\citenamefont
  {Brilliantov}, \citenamefont {Krapivsky}, \citenamefont {Bodrova},
  \citenamefont {Spahn}, \citenamefont {Hayakawa}, \citenamefont {Stadnichuk},\
  and\ \citenamefont {Schmidt}}]{brilliantov2015size}%
  \BibitemOpen
  \bibfield  {author} {\bibinfo {author} {\bibfnamefont {Nikolai}\ \bibnamefont
  {Brilliantov}}, \bibinfo {author} {\bibfnamefont {PL}~\bibnamefont
  {Krapivsky}}, \bibinfo {author} {\bibfnamefont {Anna}\ \bibnamefont
  {Bodrova}}, \bibinfo {author} {\bibfnamefont {Frank}\ \bibnamefont {Spahn}},
  \bibinfo {author} {\bibfnamefont {Hisao}\ \bibnamefont {Hayakawa}}, \bibinfo
  {author} {\bibfnamefont {Vladimir}\ \bibnamefont {Stadnichuk}}, \ and\
  \bibinfo {author} {\bibfnamefont {J{\"u}rgen}\ \bibnamefont {Schmidt}},\
  }\bibfield  {title} {\enquote {\bibinfo {title} {Size distribution of
  particles in saturn’s rings from aggregation and fragmentation},}\
  }\href@noop {} {\bibfield  {journal} {\bibinfo  {journal} {Proceedings of the
  National Academy of Sciences USA}\ }\textbf {\bibinfo {volume} {112}},\
  \bibinfo {pages} {9536--9541} (\bibinfo {year} {2015})}\BibitemShut {NoStop}%
\bibitem [{\citenamefont {Blum}(2006)}]{blum2006dust}%
  \BibitemOpen
  \bibfield  {author} {\bibinfo {author} {\bibfnamefont {J{\"u}rgen}\
  \bibnamefont {Blum}},\ }\bibfield  {title} {\enquote {\bibinfo {title} {Dust
  agglomeration},}\ }\href@noop {} {\bibfield  {journal} {\bibinfo  {journal}
  {Advances in Physics}\ }\textbf {\bibinfo {volume} {55}},\ \bibinfo {pages}
  {881--947} (\bibinfo {year} {2006})}\BibitemShut {NoStop}%
\bibitem [{\citenamefont {Jungmann}\ \emph {et~al.}(2018)\citenamefont
  {Jungmann}, \citenamefont {Steinpilz}, \citenamefont {Teiser},\ and\
  \citenamefont {Wurm}}]{jungmann2018sticking}%
  \BibitemOpen
  \bibfield  {author} {\bibinfo {author} {\bibfnamefont {Felix}\ \bibnamefont
  {Jungmann}}, \bibinfo {author} {\bibfnamefont {Tobias}\ \bibnamefont
  {Steinpilz}}, \bibinfo {author} {\bibfnamefont {Jens}\ \bibnamefont
  {Teiser}}, \ and\ \bibinfo {author} {\bibfnamefont {Gerhard}\ \bibnamefont
  {Wurm}},\ }\bibfield  {title} {\enquote {\bibinfo {title} {Sticking and
  restitution in collisions of charged sub-mm dielectric grains},}\ }\href@noop
  {} {\bibfield  {journal} {\bibinfo  {journal} {Journal of Physics
  Communications}\ }\textbf {\bibinfo {volume} {2}},\ \bibinfo {pages} {095009}
  (\bibinfo {year} {2018})}\BibitemShut {NoStop}%
\bibitem [{\citenamefont {Lee}\ \emph {et~al.}(2015)\citenamefont {Lee},
  \citenamefont {Waitukaitis}, \citenamefont {Miskin},\ and\ \citenamefont
  {Jaeger}}]{lee2015direct}%
  \BibitemOpen
  \bibfield  {author} {\bibinfo {author} {\bibfnamefont {Victor}\ \bibnamefont
  {Lee}}, \bibinfo {author} {\bibfnamefont {Scott~R}\ \bibnamefont
  {Waitukaitis}}, \bibinfo {author} {\bibfnamefont {Marc~Z}\ \bibnamefont
  {Miskin}}, \ and\ \bibinfo {author} {\bibfnamefont {Heinrich~M}\ \bibnamefont
  {Jaeger}},\ }\bibfield  {title} {\enquote {\bibinfo {title} {Direct
  observation of particle interactions and clustering in charged granular
  streams},}\ }\href@noop {} {\bibfield  {journal} {\bibinfo  {journal} {Nature
  Physics}\ }\textbf {\bibinfo {volume} {11}},\ \bibinfo {pages} {733}
  (\bibinfo {year} {2015})}\BibitemShut {NoStop}%
\bibitem [{\citenamefont {Yoshimatsu}\ \emph {et~al.}(2017)\citenamefont
  {Yoshimatsu}, \citenamefont {Ara{\'u}jo}, \citenamefont {Wurm}, \citenamefont
  {Herrmann},\ and\ \citenamefont {Shinbrot}}]{yoshimatsu2017self}%
  \BibitemOpen
  \bibfield  {author} {\bibinfo {author} {\bibfnamefont {Ryuta}\ \bibnamefont
  {Yoshimatsu}}, \bibinfo {author} {\bibfnamefont {NAM}\ \bibnamefont
  {Ara{\'u}jo}}, \bibinfo {author} {\bibfnamefont {Gerhard}\ \bibnamefont
  {Wurm}}, \bibinfo {author} {\bibfnamefont {Hans~J}\ \bibnamefont {Herrmann}},
  \ and\ \bibinfo {author} {\bibfnamefont {Troy}\ \bibnamefont {Shinbrot}},\
  }\bibfield  {title} {\enquote {\bibinfo {title} {Self-charging of identical
  grains in the absence of an external field},}\ }\href@noop {} {\bibfield
  {journal} {\bibinfo  {journal} {Scientific Reports}\ }\textbf {\bibinfo
  {volume} {7}},\ \bibinfo {pages} {39996} (\bibinfo {year}
  {2017})}\BibitemShut {NoStop}%
\bibitem [{\citenamefont {Poppe}\ \emph {et~al.}(2000)\citenamefont {Poppe},
  \citenamefont {Blum},\ and\ \citenamefont {Henning}}]{poppe2000experiments}%
  \BibitemOpen
  \bibfield  {author} {\bibinfo {author} {\bibfnamefont {Torsten}\ \bibnamefont
  {Poppe}}, \bibinfo {author} {\bibfnamefont {J{\"u}rgen}\ \bibnamefont
  {Blum}}, \ and\ \bibinfo {author} {\bibfnamefont {Thomas}\ \bibnamefont
  {Henning}},\ }\bibfield  {title} {\enquote {\bibinfo {title} {Experiments on
  collisional grain charging of micron-sized preplanetary dust},}\ }\href@noop
  {} {\bibfield  {journal} {\bibinfo  {journal} {The Astrophysical Journal}\
  }\textbf {\bibinfo {volume} {533}},\ \bibinfo {pages} {472} (\bibinfo {year}
  {2000})}\BibitemShut {NoStop}%
\bibitem [{\citenamefont {Haeberle}\ \emph {et~al.}(2018)\citenamefont
  {Haeberle}, \citenamefont {Schella}, \citenamefont {Sperl}, \citenamefont
  {Schr{\"o}ter},\ and\ \citenamefont {Born}}]{haeberle2018double}%
  \BibitemOpen
  \bibfield  {author} {\bibinfo {author} {\bibfnamefont {Jan}\ \bibnamefont
  {Haeberle}}, \bibinfo {author} {\bibfnamefont {Andr{\'e}}\ \bibnamefont
  {Schella}}, \bibinfo {author} {\bibfnamefont {Matthias}\ \bibnamefont
  {Sperl}}, \bibinfo {author} {\bibfnamefont {Matthias}\ \bibnamefont
  {Schr{\"o}ter}}, \ and\ \bibinfo {author} {\bibfnamefont {Philip}\
  \bibnamefont {Born}},\ }\bibfield  {title} {\enquote {\bibinfo {title}
  {Double origin of stochastic granular tribocharging},}\ }\href@noop {}
  {\bibfield  {journal} {\bibinfo  {journal} {Soft matter}\ } (\bibinfo {year}
  {2018})}\BibitemShut {NoStop}%
\bibitem [{\citenamefont {Spahn}\ and\ \citenamefont
  {Sei{\ss}}(2015)}]{spahn2015granular}%
  \BibitemOpen
  \bibfield  {author} {\bibinfo {author} {\bibfnamefont {Frank}\ \bibnamefont
  {Spahn}}\ and\ \bibinfo {author} {\bibfnamefont {Martin}\ \bibnamefont
  {Sei{\ss}}},\ }\bibfield  {title} {\enquote {\bibinfo {title} {Granular
  matter: Charges dropped},}\ }\href@noop {} {\bibfield  {journal} {\bibinfo
  {journal} {Nature Physics}\ }\textbf {\bibinfo {volume} {11}},\ \bibinfo
  {pages} {709} (\bibinfo {year} {2015})}\BibitemShut {NoStop}%
\bibitem [{\citenamefont {Ivlev}\ \emph {et~al.}(2002)\citenamefont {Ivlev},
  \citenamefont {Morfill},\ and\ \citenamefont
  {Konopka}}]{ivlev2002coagulation}%
  \BibitemOpen
  \bibfield  {author} {\bibinfo {author} {\bibfnamefont {AV}~\bibnamefont
  {Ivlev}}, \bibinfo {author} {\bibfnamefont {GE}~\bibnamefont {Morfill}}, \
  and\ \bibinfo {author} {\bibfnamefont {U}~\bibnamefont {Konopka}},\
  }\bibfield  {title} {\enquote {\bibinfo {title} {Coagulation of charged
  microparticles in neutral gas and charge-induced gel transitions},}\
  }\href@noop {} {\bibfield  {journal} {\bibinfo  {journal} {Physical Review
  Letters}\ }\textbf {\bibinfo {volume} {89}},\ \bibinfo {pages} {195502}
  (\bibinfo {year} {2002})}\BibitemShut {NoStop}%
\bibitem [{\citenamefont {Dammer}\ and\ \citenamefont
  {Wolf}(2004)}]{dammer2004self}%
  \BibitemOpen
  \bibfield  {author} {\bibinfo {author} {\bibfnamefont {Stephan~M}\
  \bibnamefont {Dammer}}\ and\ \bibinfo {author} {\bibfnamefont {Dietrich~E}\
  \bibnamefont {Wolf}},\ }\bibfield  {title} {\enquote {\bibinfo {title}
  {Self-focusing dynamics in monopolarly charged suspensions},}\ }\href@noop {}
  {\bibfield  {journal} {\bibinfo  {journal} {Physical Review Lletters}\
  }\textbf {\bibinfo {volume} {93}},\ \bibinfo {pages} {150602} (\bibinfo
  {year} {2004})}\BibitemShut {NoStop}%
\bibitem [{\citenamefont {M{\"u}ller}\ and\ \citenamefont
  {Luding}(2011)}]{muller2011homogeneous}%
  \BibitemOpen
  \bibfield  {author} {\bibinfo {author} {\bibfnamefont {Micha-Klaus}\
  \bibnamefont {M{\"u}ller}}\ and\ \bibinfo {author} {\bibfnamefont {Stefan}\
  \bibnamefont {Luding}},\ }\bibfield  {title} {\enquote {\bibinfo {title}
  {Homogeneous cooling with repulsive and attractive long-range potentials},}\
  }\href@noop {} {\bibfield  {journal} {\bibinfo  {journal} {Mathematical
  Modelling of Natural Phenomena}\ }\textbf {\bibinfo {volume} {6}},\ \bibinfo
  {pages} {118--150} (\bibinfo {year} {2011})}\BibitemShut {NoStop}%
\bibitem [{\citenamefont {Ulrich}\ \emph {et~al.}(2009)\citenamefont {Ulrich},
  \citenamefont {Aspelmeier}, \citenamefont {Roeller}, \citenamefont
  {Fingerle}, \citenamefont {Herminghaus},\ and\ \citenamefont
  {Zippelius}}]{ulrich2009cooling}%
  \BibitemOpen
  \bibfield  {author} {\bibinfo {author} {\bibfnamefont {Stephan}\ \bibnamefont
  {Ulrich}}, \bibinfo {author} {\bibfnamefont {Timo}\ \bibnamefont
  {Aspelmeier}}, \bibinfo {author} {\bibfnamefont {Klaus}\ \bibnamefont
  {Roeller}}, \bibinfo {author} {\bibfnamefont {Axel}\ \bibnamefont
  {Fingerle}}, \bibinfo {author} {\bibfnamefont {Stephan}\ \bibnamefont
  {Herminghaus}}, \ and\ \bibinfo {author} {\bibfnamefont {Annette}\
  \bibnamefont {Zippelius}},\ }\bibfield  {title} {\enquote {\bibinfo {title}
  {Cooling and aggregation in wet granulates},}\ }\href@noop {} {\bibfield
  {journal} {\bibinfo  {journal} {Physical Review Letters}\ }\textbf {\bibinfo
  {volume} {102}},\ \bibinfo {pages} {148002} (\bibinfo {year}
  {2009})}\BibitemShut {NoStop}%
\bibitem [{\citenamefont {Brilliantov}\ \emph {et~al.}(2018)\citenamefont
  {Brilliantov}, \citenamefont {Formella},\ and\ \citenamefont
  {P{\"o}schel}}]{brilliantov2018increasing}%
  \BibitemOpen
  \bibfield  {author} {\bibinfo {author} {\bibfnamefont {Nikolai~V}\
  \bibnamefont {Brilliantov}}, \bibinfo {author} {\bibfnamefont {Arno}\
  \bibnamefont {Formella}}, \ and\ \bibinfo {author} {\bibfnamefont {Thorsten}\
  \bibnamefont {P{\"o}schel}},\ }\bibfield  {title} {\enquote {\bibinfo {title}
  {Increasing temperature of cooling granular gases},}\ }\href@noop {}
  {\bibfield  {journal} {\bibinfo  {journal} {Nature communications}\ }\textbf
  {\bibinfo {volume} {9}},\ \bibinfo {pages} {797} (\bibinfo {year}
  {2018})}\BibitemShut {NoStop}%
\bibitem [{\citenamefont {Liu}\ and\ \citenamefont
  {Hrenya}(2018)}]{liu2018cluster}%
  \BibitemOpen
  \bibfield  {author} {\bibinfo {author} {\bibfnamefont {Peiyuan}\ \bibnamefont
  {Liu}}\ and\ \bibinfo {author} {\bibfnamefont {Christine~M.}\ \bibnamefont
  {Hrenya}},\ }\bibfield  {title} {\enquote {\bibinfo {title} {Cluster-induced
  deagglomeration in dilute gravity-driven gas-solid flows of cohesive
  grains},}\ }\href@noop {} {\bibfield  {journal} {\bibinfo  {journal} {Phys.
  Rev. Lett.}\ }\textbf {\bibinfo {volume} {121}},\ \bibinfo {pages} {238001}
  (\bibinfo {year} {2018})}\BibitemShut {NoStop}%
\bibitem [{\citenamefont {Takada}\ \emph {et~al.}(2017)\citenamefont {Takada},
  \citenamefont {Serero},\ and\ \citenamefont
  {P{\"o}schel}}]{takada2017homogeneous}%
  \BibitemOpen
  \bibfield  {author} {\bibinfo {author} {\bibfnamefont {Satoshi}\ \bibnamefont
  {Takada}}, \bibinfo {author} {\bibfnamefont {Dan}\ \bibnamefont {Serero}}, \
  and\ \bibinfo {author} {\bibfnamefont {Thorsten}\ \bibnamefont
  {P{\"o}schel}},\ }\bibfield  {title} {\enquote {\bibinfo {title} {Homogeneous
  cooling state of dilute granular gases of charged particles},}\ }\href@noop
  {} {\bibfield  {journal} {\bibinfo  {journal} {Physics of Fluids}\ }\textbf
  {\bibinfo {volume} {29}},\ \bibinfo {pages} {083303} (\bibinfo {year}
  {2017})}\BibitemShut {NoStop}%
\bibitem [{\citenamefont {Kolehmainen}\ \emph {et~al.}(2018)\citenamefont
  {Kolehmainen}, \citenamefont {Ozel}, \citenamefont {Gu}, \citenamefont
  {Shinbrot},\ and\ \citenamefont {Sundaresan}}]{shinbrot2018parlad}%
  \BibitemOpen
  \bibfield  {author} {\bibinfo {author} {\bibfnamefont {Jari}\ \bibnamefont
  {Kolehmainen}}, \bibinfo {author} {\bibfnamefont {Ali}\ \bibnamefont {Ozel}},
  \bibinfo {author} {\bibfnamefont {Yile}\ \bibnamefont {Gu}}, \bibinfo
  {author} {\bibfnamefont {Troy}\ \bibnamefont {Shinbrot}}, \ and\ \bibinfo
  {author} {\bibfnamefont {Sankaran}\ \bibnamefont {Sundaresan}},\ }\bibfield
  {title} {\enquote {\bibinfo {title} {Effects of polarization on
  particle-laden flows},}\ }\href@noop {} {\bibfield  {journal} {\bibinfo
  {journal} {Phys. Rev. Lett.}\ }\textbf {\bibinfo {volume} {121}},\ \bibinfo
  {pages} {124503} (\bibinfo {year} {2018})}\BibitemShut {NoStop}%
\bibitem [{\citenamefont {Singh}\ and\ \citenamefont
  {Mazza}(2018)}]{singh2018early}%
  \BibitemOpen
  \bibfield  {author} {\bibinfo {author} {\bibfnamefont {Chamkor}\ \bibnamefont
  {Singh}}\ and\ \bibinfo {author} {\bibfnamefont {Marco~G}\ \bibnamefont
  {Mazza}},\ }\bibfield  {title} {\enquote {\bibinfo {title} {Early-stage
  aggregation in three-dimensional charged granular gas},}\ }\href@noop {}
  {\bibfield  {journal} {\bibinfo  {journal} {Physical Review E}\ }\textbf
  {\bibinfo {volume} {97}},\ \bibinfo {pages} {022904} (\bibinfo {year}
  {2018})}\BibitemShut {NoStop}%
\bibitem [{\citenamefont {Pitaevskii}\ and\ \citenamefont
  {Lifshitz}(2012)}]{pitaevskii2012physical}%
  \BibitemOpen
  \bibfield  {author} {\bibinfo {author} {\bibfnamefont {LP}~\bibnamefont
  {Pitaevskii}}\ and\ \bibinfo {author} {\bibfnamefont {EM}~\bibnamefont
  {Lifshitz}},\ }\href@noop {} {\emph {\bibinfo {title} {Physical kinetics}}},\
  Vol.~\bibinfo {volume} {10}\ (\bibinfo  {publisher} {Butterworth-Heinemann},\
  \bibinfo {year} {2012})\BibitemShut {NoStop}%
\bibitem [{\citenamefont {Brilliantov}\ and\ \citenamefont
  {P{\"o}schel}(2010)}]{brilliantov2010kinetic}%
  \BibitemOpen
  \bibfield  {author} {\bibinfo {author} {\bibfnamefont {Nikolai~V}\
  \bibnamefont {Brilliantov}}\ and\ \bibinfo {author} {\bibfnamefont
  {Thorsten}\ \bibnamefont {P{\"o}schel}},\ }\href@noop {} {\emph {\bibinfo
  {title} {Kinetic theory of granular gases}}}\ (\bibinfo  {publisher} {Oxford
  University Press},\ \bibinfo {year} {2010})\BibitemShut {NoStop}%
\bibitem [{\citenamefont {Smirnov}(1990)}]{smirnov1990properties}%
  \BibitemOpen
  \bibfield  {author} {\bibinfo {author} {\bibfnamefont
  {Boris~Mikha{\u\i}lovich}\ \bibnamefont {Smirnov}},\ }\bibfield  {title}
  {\enquote {\bibinfo {title} {The properties of fractal clusters},}\
  }\href@noop {} {\bibfield  {journal} {\bibinfo  {journal} {Physics Reports}\
  }\textbf {\bibinfo {volume} {188}},\ \bibinfo {pages} {1--78} (\bibinfo
  {year} {1990})}\BibitemShut {NoStop}%
\bibitem [{\citenamefont {Mandelbrot}(1977)}]{mandelbrot1977fractals}%
  \BibitemOpen
  \bibfield  {author} {\bibinfo {author} {\bibfnamefont {Benoit}\ \bibnamefont
  {Mandelbrot}},\ }\href@noop {} {\emph {\bibinfo {title} {Fractals}}}\
  (\bibinfo  {publisher} {Freeman San Francisco},\ \bibinfo {year}
  {1977})\BibitemShut {NoStop}%
\bibitem [{\citenamefont {Jullien}(1987)}]{jullien1987aggregation}%
  \BibitemOpen
  \bibfield  {author} {\bibinfo {author} {\bibfnamefont {R}~\bibnamefont
  {Jullien}},\ }\bibfield  {title} {\enquote {\bibinfo {title} {Aggregation
  phenomena and fractal aggregates},}\ }\href@noop {} {\bibfield  {journal}
  {\bibinfo  {journal} {Contemporary Physics}\ }\textbf {\bibinfo {volume}
  {28}},\ \bibinfo {pages} {477--493} (\bibinfo {year} {1987})}\BibitemShut
  {NoStop}%
\bibitem [{\citenamefont {Lebovka}(2012)}]{lebovka2012aggregation}%
  \BibitemOpen
  \bibfield  {author} {\bibinfo {author} {\bibfnamefont {Nikolai~I}\
  \bibnamefont {Lebovka}},\ }\bibfield  {title} {\enquote {\bibinfo {title}
  {Aggregation of charged colloidal particles},}\ }in\ \href@noop {} {\emph
  {\bibinfo {booktitle} {Polyelectrolyte Complexes in the Dispersed and Solid
  State I}}}\ (\bibinfo  {publisher} {Springer},\ \bibinfo {year} {2012})\ pp.\
  \bibinfo {pages} {57--96}\BibitemShut {NoStop}%
\bibitem [{\citenamefont {Kempf}\ \emph {et~al.}(1999)\citenamefont {Kempf},
  \citenamefont {Pfalzner},\ and\ \citenamefont {Henning}}]{kempf1999n}%
  \BibitemOpen
  \bibfield  {author} {\bibinfo {author} {\bibfnamefont {Sascha}\ \bibnamefont
  {Kempf}}, \bibinfo {author} {\bibfnamefont {Susanne}\ \bibnamefont
  {Pfalzner}}, \ and\ \bibinfo {author} {\bibfnamefont {Thomas~K}\ \bibnamefont
  {Henning}},\ }\bibfield  {title} {\enquote {\bibinfo {title}
  {N-particle-simulations of dust growth: I. growth driven by brownian
  motion},}\ }\href@noop {} {\bibfield  {journal} {\bibinfo  {journal}
  {Icarus}\ }\textbf {\bibinfo {volume} {141}},\ \bibinfo {pages} {388--398}
  (\bibinfo {year} {1999})}\BibitemShut {NoStop}%
\bibitem [{\citenamefont {Hummel}\ \emph {et~al.}(2016)\citenamefont {Hummel},
  \citenamefont {Clewett},\ and\ \citenamefont {Mazza}}]{hummel2016universal}%
  \BibitemOpen
  \bibfield  {author} {\bibinfo {author} {\bibfnamefont {Mathias}\ \bibnamefont
  {Hummel}}, \bibinfo {author} {\bibfnamefont {James~PD}\ \bibnamefont
  {Clewett}}, \ and\ \bibinfo {author} {\bibfnamefont {Marco~G}\ \bibnamefont
  {Mazza}},\ }\bibfield  {title} {\enquote {\bibinfo {title} {A universal
  scaling law for the evolution of granular gases},}\ }\href@noop {} {\bibfield
   {journal} {\bibinfo  {journal} {EPL (Europhysics Letters)}\ }\textbf
  {\bibinfo {volume} {114}},\ \bibinfo {pages} {10002} (\bibinfo {year}
  {2016})}\BibitemShut {NoStop}%
\bibitem [{\citenamefont {Lee}\ \emph {et~al.}(2018)\citenamefont {Lee},
  \citenamefont {James}, \citenamefont {Waitukaitis},\ and\ \citenamefont
  {Jaeger}}]{lee2018collisional}%
  \BibitemOpen
  \bibfield  {author} {\bibinfo {author} {\bibfnamefont {Victor}\ \bibnamefont
  {Lee}}, \bibinfo {author} {\bibfnamefont {Nicole~M}\ \bibnamefont {James}},
  \bibinfo {author} {\bibfnamefont {Scott~R}\ \bibnamefont {Waitukaitis}}, \
  and\ \bibinfo {author} {\bibfnamefont {Heinrich~M}\ \bibnamefont {Jaeger}},\
  }\bibfield  {title} {\enquote {\bibinfo {title} {Collisional charging of
  individual submillimeter particles: Using ultrasonic levitation to initiate
  and track charge transfer},}\ }\href@noop {} {\bibfield  {journal} {\bibinfo
  {journal} {Physical Review Materials}\ }\textbf {\bibinfo {volume} {2}},\
  \bibinfo {pages} {035602} (\bibinfo {year} {2018})}\BibitemShut {NoStop}%
\bibitem [{\citenamefont {Zhang}\ \emph {et~al.}(2015)\citenamefont {Zhang},
  \citenamefont {P{\"a}htz}, \citenamefont {Liu}, \citenamefont {Wang},
  \citenamefont {Zhang}, \citenamefont {Shen}, \citenamefont {Ji},\ and\
  \citenamefont {Cai}}]{zhang2015electric}%
  \BibitemOpen
  \bibfield  {author} {\bibinfo {author} {\bibfnamefont {Yanzhen}\ \bibnamefont
  {Zhang}}, \bibinfo {author} {\bibfnamefont {Thomas}\ \bibnamefont
  {P{\"a}htz}}, \bibinfo {author} {\bibfnamefont {Yonghong}\ \bibnamefont
  {Liu}}, \bibinfo {author} {\bibfnamefont {Xiaolong}\ \bibnamefont {Wang}},
  \bibinfo {author} {\bibfnamefont {Rui}\ \bibnamefont {Zhang}}, \bibinfo
  {author} {\bibfnamefont {Yang}\ \bibnamefont {Shen}}, \bibinfo {author}
  {\bibfnamefont {Renjie}\ \bibnamefont {Ji}}, \ and\ \bibinfo {author}
  {\bibfnamefont {Baoping}\ \bibnamefont {Cai}},\ }\bibfield  {title} {\enquote
  {\bibinfo {title} {Electric field and humidity trigger contact
  electrification},}\ }\href@noop {} {\bibfield  {journal} {\bibinfo  {journal}
  {Physical Review X}\ }\textbf {\bibinfo {volume} {5}},\ \bibinfo {pages}
  {011002} (\bibinfo {year} {2015})}\BibitemShut {NoStop}%
\bibitem [{\citenamefont {Yan}\ \emph {et~al.}(2016)\citenamefont {Yan},
  \citenamefont {Han}, \citenamefont {Zhang}, \citenamefont {Xu}, \citenamefont
  {Luijten},\ and\ \citenamefont {Granick}}]{yan2016reconfiguring}%
  \BibitemOpen
  \bibfield  {author} {\bibinfo {author} {\bibfnamefont {Jing}\ \bibnamefont
  {Yan}}, \bibinfo {author} {\bibfnamefont {Ming}\ \bibnamefont {Han}},
  \bibinfo {author} {\bibfnamefont {Jie}\ \bibnamefont {Zhang}}, \bibinfo
  {author} {\bibfnamefont {Cong}\ \bibnamefont {Xu}}, \bibinfo {author}
  {\bibfnamefont {Erik}\ \bibnamefont {Luijten}}, \ and\ \bibinfo {author}
  {\bibfnamefont {Steve}\ \bibnamefont {Granick}},\ }\bibfield  {title}
  {\enquote {\bibinfo {title} {Reconfiguring active particles by electrostatic
  imbalance},}\ }\href@noop {} {\bibfield  {journal} {\bibinfo  {journal}
  {Nature Materials}\ }\textbf {\bibinfo {volume} {15}},\ \bibinfo {pages}
  {1095} (\bibinfo {year} {2016})}\BibitemShut {NoStop}%
\bibitem [{\citenamefont {Friedl}\ and\ \citenamefont
  {Gilmour}(2009)}]{friedl2009collective}%
  \BibitemOpen
  \bibfield  {author} {\bibinfo {author} {\bibfnamefont {Peter}\ \bibnamefont
  {Friedl}}\ and\ \bibinfo {author} {\bibfnamefont {Darren}\ \bibnamefont
  {Gilmour}},\ }\bibfield  {title} {\enquote {\bibinfo {title} {Collective cell
  migration in morphogenesis, regeneration and cancer},}\ }\href@noop {}
  {\bibfield  {journal} {\bibinfo  {journal} {Nature Reviews Molecular Cell
  Biology}\ }\textbf {\bibinfo {volume} {10}},\ \bibinfo {pages} {445}
  (\bibinfo {year} {2009})}\BibitemShut {NoStop}%
\bibitem [{\citenamefont {Dash}\ \emph {et~al.}(2001)\citenamefont {Dash},
  \citenamefont {Mason},\ and\ \citenamefont {Wettlaufer}}]{dash2001theory}%
  \BibitemOpen
  \bibfield  {author} {\bibinfo {author} {\bibfnamefont {JG}~\bibnamefont
  {Dash}}, \bibinfo {author} {\bibfnamefont {BL}~\bibnamefont {Mason}}, \ and\
  \bibinfo {author} {\bibfnamefont {JS}~\bibnamefont {Wettlaufer}},\ }\bibfield
   {title} {\enquote {\bibinfo {title} {Theory of charge and mass transfer in
  ice-ice collisions},}\ }\href@noop {} {\bibfield  {journal} {\bibinfo
  {journal} {Journal of Geophysical Research: Atmospheres}\ }\textbf {\bibinfo
  {volume} {106}},\ \bibinfo {pages} {20395--20402} (\bibinfo {year}
  {2001})}\BibitemShut {NoStop}%
\bibitem [{\citenamefont {Zsom}\ \emph {et~al.}(2010)\citenamefont {Zsom},
  \citenamefont {Ormel}, \citenamefont {G{\"u}ttler}, \citenamefont {Blum},\
  and\ \citenamefont {Dullemond}}]{zsom2010outcome}%
  \BibitemOpen
  \bibfield  {author} {\bibinfo {author} {\bibfnamefont {Andras}\ \bibnamefont
  {Zsom}}, \bibinfo {author} {\bibfnamefont {CW}~\bibnamefont {Ormel}},
  \bibinfo {author} {\bibfnamefont {C}~\bibnamefont {G{\"u}ttler}}, \bibinfo
  {author} {\bibfnamefont {J}~\bibnamefont {Blum}}, \ and\ \bibinfo {author}
  {\bibfnamefont {CP}~\bibnamefont {Dullemond}},\ }\bibfield  {title} {\enquote
  {\bibinfo {title} {The outcome of protoplanetary dust growth: pebbles,
  boulders, or planetesimals?-ii. introducing the bouncing barrier},}\
  }\href@noop {} {\bibfield  {journal} {\bibinfo  {journal} {Astronomy \&
  Astrophysics}\ }\textbf {\bibinfo {volume} {513}},\ \bibinfo {pages} {A57}
  (\bibinfo {year} {2010})}\BibitemShut {NoStop}%
\end{thebibliography}
%

%
%
%
\end{document}